\documentclass[nohyper]{JHEP3}

\usepackage{amsmath,amssymb}
\usepackage[pdftex]{graphicx}
\usepackage{afterpage}
\usepackage{bbm}

\newcommand{\ie}{\textit{i.e.}}
\newcommand{\eg}{\textit{e.g.}}
\newcommand{\LT}{\left}
\newcommand{\RT}{\right}

\newcommand{\eVq}{\text{eV}^2}
\newcommand{\GeV}{\text{GeV}}
\newcommand{\cst}{\text{cst}}
\newcommand{\Dmq}{\Delta m^2}
\newcommand{\diag}{\mathop{\mathrm{diag}}}
\renewcommand{\Re}{\mathop{\mathrm{Re}}}
\renewcommand{\Im}{\mathop{\mathrm{Im}}}

\newenvironment{myitemize}%
{\begin{list}{\textbullet}{%
    \setlength{\leftmargin}{\parindent}%
    \setlength{\parsep}{0pt}%
    \setlength{\topsep}{\itemsep}%
    \setlength{\parskip}{0pt}%
    }}%
{\end{list}}

\newcommand{\PAGEFIGURE}[1]{\FIGURE[!p]{#1}\afterpage\clearpage}

\title{Neutrino oscillograms of the Earth: effects of 1-2 mixing and
  CP-violation}

\author{Evgeny Kh.~Akhmedov\\
  Max-Planck-Institut f\"ur Kernphysik, Postfach 103980, D-69029
  Heidelberg, Germany\\
  {\rm and:} National Research Centre Kurchatov Institute, Moscow,
  Russia\\
  E-mail: \email{akhmedov@mpi-hd.mpg.de}}

\author{Michele Maltoni\\
  Departamento de F\'isica Te\'orica \& Instituto de F\'isica
  Te\'orica UAM/CSIC, Facultad de Ciencias C-XI, Universidad
  Aut\'onoma de Madrid, Cantoblanco, E-28049 Madrid, Spain\\
  E-mail: \email{michele.maltoni@uam.es}}

\author{Alexei Yu.~Smirnov\\
  The Abdus Salam International Centre for Theoretical Physics,
  Strada Costiera 11, I-34014 Trieste, Italy\\
  {\rm and:} Institute for Nuclear Research, Russian Academy of
  Sciences, Moscow, Russia\\
  E-mail: \email{smirnov@ictp.trieste.it}}

\abstract{%
  We develop a comprehensive description of three flavor neutrino
  oscillations inside the Earth in terms of neutrino oscillograms in
  the whole range of nadir angles and for energies above $0.1~\GeV$.
  The effects of the 1-2 mass splitting and mixing as well the
  interference of the 1-2 and 1-3 modes of oscillations are
  quantified.  The 1-2 mass splitting and mixing lead to the
  appearance, apart from the resonance MSW peaks, of the parametric
  resonance peak for core-crossing trajectories at $E_\nu \sim 0.2$
  GeV.  We show that the interference effects, in particular CP
  violation, have a domain structure with borders determined by the
  solar and atmospheric magic lines and the lines of the interference
  phase condition.  The dependence of the oscillograms on the Dirac
  CP-violating phase is studied.  We show that for $\sin^2 2
  \theta_{13} < 0.1$ the strongest dependence of the oscillograms on
  $\delta$ is in the 1-2 and 1-3 resonance regions.}

\preprint{IFT-UAM/CSIC-08-18}


\begin{document}


\section{Introduction}
\label{sec:intro}

Substantial future progress in neutrino physics will be related to the
long baseline experiments as well as studies of the cosmic and
atmospheric neutrinos. These studies are expected to fill some of the
outstanding gaps in our knowledge of neutrino properties, such as the
value of the leptonic mixing angle $\theta_{13}$, the type of the
neutrino mass hierarchy, the octant of the mixing angle $\theta_{23}$
and the size of the Dirac-type leptonic CP-violation. They are also
expected to improve the accuracy of the determination of the already
known parameters, such as the mass squared differences $\Dmq_{21}$ and
$|\Dmq_{31}|$ and the mixing parameters $\theta_{12}$ and $\sin^2
2\theta_{23}$.

The key element of these experiments is that neutrinos propagate long
distances inside the Earth before reaching detectors, and therefore a
careful analysis of the Earth matter effects on neutrino oscillations
is necessary (see, \eg, references in~\cite{Akhmedov:2006hb,
GonzalezGarcia:2007ib} as well as Refs.~\cite{Lisi:1997yc, Liu:1997yb,
Liu:1998nb, DeRujula:1998hd, Freund:1999gy, Freund:1999vc,
Akhmedov:2000cs, Mocioiu:2000st, Akhmedov:2001kd, Banuls:2001zn,
Indumathi:2004kd, PalomaresRuiz:2004tk, Gandhi:2004bj, Gandhi:2004md,
Kimura:2004vh, Lin:2005pi, Akhmedov:2005yj, Ioannisian:2008ve,
Supanitsky:2008eq}). In a previous publication~\cite{Akhmedov:2006hb}
we have studied these matter effects by making use of ``neutrino
oscillograms'' of the Earth as the main tool. These are the contours
of constant oscillation probabilities in the plane of neutrino energy
and nadir angle.
The plots of this type were introduced by P.~Lipari in 1998
(unpublished) and then appeared in several
publications~\cite{Chizhov:1998ug, Ohlsson:1999um, Jacobson:2003wc,
Kajita:2004ga, Akhmedov:2006hb}.
The oscillograms exhibit a very rich structure with local and global
maxima and minima, including the MSW~\cite{Wolfenstein:1977ue,
Mikheev:1986gs} resonance maxima in the mantle and core of the Earth
and the parametric enhancement~\cite{Ermilova:1986xx, Akhmedov:1988kd,
Liu:1997yb, Liu:1998nb, Petcov:1998su, Akhmedov:1998ui,
Akhmedov:1998xq, Chizhov:1998ug, Chizhov:1999he} ridges for
core-crossing neutrino trajectories. It was shown
in~\cite{Akhmedov:2006hb} that all these features, including the local
and global minima and maxima as well as saddle points can be
understood in terms of various realizations of just two conditions:
the generalized amplitude and phase conditions. We refer the reader to
Ref.~\cite{Akhmedov:2006hb} for details. It has been shown that these
oscillograms are very useful for gaining an insight into the physics
of neutrino oscillations in the Earth and should help plan the future
experiments as well as interpret their data.

The analysis in~\cite{Akhmedov:2006hb} was performed in the limit of
vanishing ``solar'' mass squared splitting $\Dmq_{21}$. While this
approximation is quantitatively well justified at relatively high
neutrino energies ($E_\nu\gtrsim 3~\GeV$), it is less satisfactory at
lower energies and also misses some important 3-flavor features of
neutrino oscillations, most notably CP violation. In the present paper
we extend the study of~\cite{Akhmedov:2006hb} to the case $\Dmq_{21}
\ne 0$ and consider 3-flavor effects in neutrino oscillations in the
Earth, with the emphasis on the effects of the CP violating phase
$\delta$.
We explore in detail
\begin{myitemize}
  \item the effects of the 1-2 mixing and splitting on the oscillation
    probabilities,
    
  \item the interference of the 1-2 and 1-3 (\ie, ``solar'' and
    ``atmospheric'') amplitudes,
    
  \item the effects of and the sensitivity to the Dirac-type
    CP-violating phase $\delta$,
    
  \item the dependence of the oscillation probabilities on the neutrino
    mass hierarchy.
\end{myitemize}
We perform numerical calculations of oscillation probabilities and
also develop a simple analytic approach to interpretation of the
obtained results based on lines of three types in the neutrino
energy~--~nadir angle plane. These are the solar and atmospheric
``magic lines'', \ie, the lines on which respectively the solar or
atmospheric contributions to the transition amplitude approximately
vanish, and the interference phase lines. Construction of these curves
allows one to identify the regions in the experimental parameter space
that are most sensitive to the effects of non-vanishing phase
$\delta$. Our results can be useful for planning experiments with
atmospheric and accelerator neutrinos, as well as neutrinos of cosmic
origin.

In the present paper, as well as in~\cite{Akhmedov:2006hb}, we confine
our consideration to the study of the oscillations probabilities.
Accurate predictions for the event numbers and sensitivities of future
experiments can only be done when an information on the corresponding
detection efficiencies and systematic errors becomes available. Still,
some general statements can be made and estimates done even in the
absence of such an information; we plan to present the corresponding
analysis in a future publication.

Three-flavor effects in neutrino oscillations in the Earth have been
considered in the past~\cite{Kuo:1987km, Minakata:1998bf, Dick:1999ed,
Freund:1999gy, Ohlsson:1999um, Peres:1999yi, Minakata:1999ze,
Ota:2000hf, Parke:2000hu, Mocioiu:2001jy, Akhmedov:2001kd,
Freund:2001pn, Yokomakura:2002av, GonzalezGarcia:2002mu,
Brahmachari:2003bk, Jacobson:2003wc, Peres:2003wd, Akhmedov:2004ny,
Minakata:2004pg, Kimura:2004vh, GonzalezGarcia:2004cu,
Takamura:2005df, Kimura:2006qx, Kimura:2006hy, Liao:2007re,
Kimura:2007mu, Kimura:2008nq}.
The main new results of the present paper are the analysis of the
neutrino oscillations in the Earth in terms of the aforementioned
three sets of curves, and use of the neutrino oscillograms for a
detailed study of the domains of the parameter space that are most
sensitive to the value of CP-violating phase $\delta$.

The paper is organized as follows. In Sec.~\ref{sec:3fosc} we give
general 3-flavor expressions for the oscillation probabilities in
matter. We present exact analytic results for matter with constant
density and introduce a factorization approximation. In
Sec.~\ref{sec:12split} present the neutrino oscillograms for different
oscillation channels and discuss the effects of non-vanishing mixing
and splitting in the 1-2 sector on these oscillograms. We consider
features of the oscillograms for the inverted mass hierarchy.
Sec.~\ref{sec:cpviol} contains the discussion of the effects of the
CP-violating phase $\delta$ and their analysis in terms of the three
sets of special lines.  In this section we also discuss the
sensitivity of the oscillation probabilities to the phase $\delta$ and
its dependence on neutrino energy and nadir angle (baseline length).
Conclusions follow in Sec.~\ref{sec:concl}.


\section{Three-flavor neutrino oscillations in matter}
\label{sec:3fosc}


\subsection{Evolution matrix and probabilities for symmetric profile}

We consider mixing of the three flavor neutrinos, $\nu_f \equiv
(\nu_e, \nu_\mu, \nu_\tau)^T$. The mixing matrix $U$, defined through
$\nu_f = U \nu_m$, where $\nu_m = (\nu_1, \nu_2, \nu_3)^T$ is the
vector of neutrino mass eigenstates, can be parametrized as
\begin{equation}
    \label{eq:mixing}
    U = U_{23} \, I_\delta \, U_{13} \, I_{-\delta} \, U_{12} \,.
\end{equation}
Here the matrices $U_{ij} = U_{ij}(\theta_{ij})$ describe rotations in
the $ij$-planes by the angles $\theta_{ij}$, and $I_\delta \equiv
\diag(1, 1, e^{i\delta})$, where $\delta$ is the Dirac-type
CP-violating phase.

Evolution of the system in matter is described by the equation
\begin{equation}
    \label{eq:evolution}
    i \frac{d \nu_f}{dt} =
    \LT( \frac{U M^2 U^\dagger}{2 E_\nu} + \hat{V} \RT) \nu_f \,,
\end{equation}
where $E_\nu$ is the neutrino energy and $M^2 \equiv \diag (0,
\Dmq_{21}, \Dmq_{31})$ is the diagonal matrix of neutrino mass squared
differences with $\Dmq_{ji} \equiv m_j^2 - m_i^2$. $\hat{V} =
\diag(V_e, 0, 0)$ is the matrix of matter-induced neutrino potentials
with $V_e \equiv \sqrt{2} \, G_F N_e$, $G_F$ and $N_e$ being the Fermi
constant and the electron number density, respectively. The evolution
matrix $S(t, t_0)$ (the matrix of oscillation amplitudes) defined
through $\nu(t) = S(t, t_0) \, \nu(t_0)$ satisfies the same
Eq.~\eqref{eq:evolution} with the initial condition $S(t_0, t_0) =
\mathbbm{1}$.

It is convenient to consider the evolution of the neutrino system in
the propagation basis $\tilde{\nu} = (\nu_e, \tilde{\nu}_{2},
\tilde{\nu}_{3})^T$ defined through the relation
\begin{equation}
    \label{eq:basisrel}
    \nu_f = U_{23} \, I_\delta \, \tilde{\nu}
\end{equation}
with $\tilde{\nu} = U_{13} \, I_{-\delta} \, U_{12} \, \nu$. According
to Eqs.~\eqref{eq:evolution} and~\eqref{eq:mixing}, the Hamiltonian
$\tilde{H}$ which describes the evolution of the neutrino vector of
state $\tilde{\nu}$ is
\begin{equation}
    \tilde{H} = \frac{1}{2E_\nu} U_{13} \, U_{12} \, M^2 \,
    U^\dagger_{12} \, U^\dagger_{13} + \hat{V} \,.
\end{equation}
This Hamiltonian does not depend on the 2-3 mixing and the
CP-violating phase. The dependence on these parameters appears when
one projects the initial flavor state on the propagation basis and the
final state back onto the original flavor basis. Explicitly, the
Hamiltonian $\tilde{H}$ reads
\begin{equation}
    \label{eq:matr1}
    \tilde{H} = \frac{\Dmq_{31}}{2E_\nu}
    \begin{pmatrix}
	s_{13}^2 + s_{12}^2\, c_{13}^2\, r_\Delta + 2 V_e \, E_\nu / \Dmq_{31}
	& s_{12}\, c_{12}\, c_{13}\, r_\Delta
	& s_{13}\, c_{13}(1 - s_{12}^2\, r_\Delta)
	\\
	\dots
	& c_{12}^2\, r_\Delta
	& - s_{12}\, c_{12}\, s_{13}\, r_\Delta
	\\
	\dots
	& \dots
	& c_{13}^2 + s_{12}^2\, s_{13}^2\, r_\Delta
    \end{pmatrix}.
\end{equation}
Here $r_\Delta \equiv \Dmq_{21} / \Dmq_{31}$, $c_{ij} \equiv \cos
\theta_{ij}$ and $s_{ij} \equiv \sin \theta_{ij}$. We introduce the
evolution matrix (the matrix of transition and survival amplitudes) in
the basis $(\nu_e, \tilde{\nu}_2, \tilde{\nu}_3)$ as
\begin{equation}
    \label{eq:matr2}
    \tilde{S} =
    \begin{pmatrix}
	A_{ee} & A_{e\tilde{2}} & A_{e\tilde{3}} \\
	A_{\tilde{2}e} & A_{\tilde{2}\tilde{2}} & A_{\tilde{2}\tilde{3}} \\
	A_{\tilde{3}e} & A_{\tilde{3}\tilde{2}} & A_{\tilde{3}\tilde{3}}
    \end{pmatrix}.
\end{equation}
This matrix satisfies the Schr\"odinger equation with the Hamiltonian
$\tilde{H}$. Then, according to Eq.~\eqref{eq:basisrel}, the evolution
matrix in the flavor basis $S$ is
\begin{equation}
    \label{eq:ftild}
    S = \tilde{U} \, \tilde{S} \, \tilde{U}^{\dagger}, \qquad
    \tilde{U} \equiv U_{23} \, I_\delta.
\end{equation}
As follows immediately from the form of the Hamiltonian in
Eq.~\eqref{eq:matr1}, the amplitudes have the following hierarchy:
\begin{equation}
    A_{e\tilde{2}}, A_{\tilde{2}e} \sim r_\Delta, \qquad
    A_{e\tilde{3}}, A_{\tilde{3}e} \sim s_{13}, \qquad
    A_{\tilde{3}\tilde{2}}, A_{\tilde{2}\tilde{3}} \sim s_{13} r_\Delta,
\end{equation}
\ie, $A_{\tilde{2}\tilde{3}}$ and $A_{\tilde{3}\tilde{2}}$ are the
smallest amplitudes.

In terms of the propagation-basis amplitudes, according to
Eqs.~\eqref{eq:matr2} and \eqref{eq:ftild} (see 
also~\cite{Yokomakura:2002av}), the matrix $S$ is given by
\begin{equation}
    \label{eq:matr3}
    S =
    \begin{pmatrix}
	A_{ee}
	& c_{23} A_{e\tilde{2}} + s_{23} e^{-i\delta} A_{e\tilde{3}}
	& - s_{23} A_{e\tilde{2}} + c_{23}e^{-i\delta} A_{e\tilde{3}}
	\\
	c_{23} A_{\tilde{2}e} + s_{23} e^{i\delta} A_{\tilde{3}e}
	& c_{23}^2 A_{\tilde{2}\tilde{2}} + s_{23}^2 A_{\tilde{3}\tilde{3}} + K_{\mu\mu}
	& - s_{23} c_{23} (A_{\tilde{2}\tilde{2}} - A_{\tilde{3}\tilde{3}}) + K_{\mu\tau}
	\\
	- s_{23} A_{\tilde{2}e} + c_{23}e^{i\delta} A_{\tilde{3}e}
	& - s_{23}c_{23} (A_{\tilde{2}\tilde{2}} - A_{\tilde{3}\tilde{3}}) + K_{\tau\mu}
	& s_{23}^2 A_{\tilde{2}\tilde{2}} + c_{23}^2 A_{\tilde{3}\tilde{3}} + K_{\tau\tau}
    \end{pmatrix},
\end{equation}
where
\begin{equation}\begin{aligned}
    K_{\mu\mu}
    &\equiv s_{23} c_{23} (e^{-i\delta}
    A_{\tilde{2}\tilde{3}} + e^{i\delta} A_{\tilde{3}\tilde{2}}) \,,
    \\
    K_{\mu\tau}
    &\equiv c_{23}^2 e^{-i\delta} A_{\tilde{2}\tilde{3}}
    - s_{23}^2 e^{i\delta} A_{\tilde{3}\tilde{2}} \,,
    \\
    K_{\tau\mu}
    &= K_{\mu\tau}(\delta \to -\delta,\,
    \tilde{2} \leftrightarrow \tilde{3}) \,,
    \\
    K_{\tau\tau} &= -K_{\mu\mu} \,.
\end{aligned}\end{equation}
Notice that $K_{\alpha \beta}$ ($\alpha, \beta = \mu, \tau$) are
proportional to the small amplitudes $A_{\tilde{2}\tilde{3}}$ and
$A_{\tilde{3}\tilde{2}}$.

Since the Earth density profile is to a good approximation symmetric
with respect to the midpoint of the neutrino trajectory and there is
no fundamental CP- (and T-) violation in the propagation basis, the
neutrino evolution is T-invariant in this basis, which
yields~\cite{Akhmedov:2001kd}
\begin{equation}
    \label{eq:sym}
    A_{\tilde{2}e} = A_{e\tilde{2}} \,, \qquad
    A_{\tilde{3}e} = A_{e\tilde{3}} \,, \qquad
    A_{\tilde{3}\tilde{2}} = A_{\tilde{2}\tilde{3}} \,.
\end{equation}
Therefore for $K_{\alpha\beta}$ we obtain
\begin{equation}\begin{aligned}
    \label{eq:K}
    K_{\mu\tau}
    &= A_{\tilde{2}\tilde{3}} (\cos 2\theta_{23} \cos\delta - i \sin\delta),
    \\
    K_{\tau\mu}
    &= A_{\tilde{2}\tilde{3}} (\cos 2\theta_{23} \cos\delta + i \sin\delta),
    \\
    K_{\mu\mu} &= -K_{\tau\tau} = A_{\tilde{2}\tilde{3}}
    \sin 2\theta_{23} \cos\delta \,.
\end{aligned}\end{equation}
Notice that the diagonal elements $S_{\mu\mu}$ and $S_{\tau\tau}$ of
the evolution matrix \eqref{eq:matr3} depend on the CP phase only via
$\cos \delta$, whereas $S_{ee}$ does not depend on $\delta$ at all.
The latter is a consequence of our use of the standard parametrization
\eqref{eq:mixing} for the leptonic mixing matrix.

The oscillation probabilities are expressed through the matrix
elements of $S$ as
\begin{equation}
    \label{eq:P}
    P_{\alpha\beta} \equiv P(\nu_\alpha\to \nu_\beta)
    = |S_{\beta\alpha}|^2 \quad\text{with}\quad
    \alpha, \beta = e, \mu, \tau \,.
\end{equation}
From Eqs.~\eqref{eq:P}, \eqref{eq:matr3}, \eqref{eq:sym}
and~\eqref{eq:K} one finds for the probabilities $P_{\alpha\beta}$
\begin{align}
    \label{eq:Pmue}
    P_{\mu e} &= c_{23}^2 |A_{e\tilde{2}}|^2
    + s_{23}^2 |A_{e\tilde{3}}|^2 + 2\, s_{23}\, c_{23}\,
    \Re( e^{-i \delta} A_{e\tilde{2}}^* A_{e\tilde{3}} ) \,,
    \\
    \label{eq:Ptaue}
    P_{\tau e} &= s_{23}^2 |A_{e\tilde{2}}|^2
    + c_{23}^2 |A_{e\tilde{3}}|^2 - 2\, s_{23}\, c_{23}\,
    \Re(e^{-i\delta} A_{e\tilde{2}}^* A_{e\tilde{3}}) \,,
    \\
    \label{eq:Pmumu}
    P_{\mu\mu} &= |c_{23}^2 A_{\tilde{2}\tilde{2}}
    + s_{23}^2 A_{\tilde{3}\tilde{3}}
    + 2\, s_{23}\, c_{23}\, \cos\delta A_{\tilde{2}\tilde{3}}|^2 \,,
    \\
    \label{eq:Pmutau}
    P_{\mu\tau} &=
    |s_{23}\, c_{23} (A_{\tilde{3}\tilde{3}} - A_{\tilde{2}\tilde{2}})
    + (\cos 2\theta_{23}\, \cos\delta + i \sin\delta)
    A_{\tilde{2}\tilde{3}}|^2 \,.
\end{align}
The probabilities $P_{\beta\alpha}$ are obtained from
$P_{\alpha\beta}$ through the substitution $\delta \to
-\delta$:\footnote{Note that in a matter with an asymmetric density
profile one would also have to substitute $V\to \tilde{V}$, where
$\tilde{V}$ is the reverse profile corresponding to the interchanged
positions of the neutrino source and detector~\cite{Akhmedov:2001kd}).
This, in particular, means that one would have to distinguish $A_{ij}$
from $A_{ji}$ and $K_{\beta\alpha}$ from $K_{\alpha\beta}$.}
\begin{equation}
    \label{eq:Trev}
    P_{\beta\alpha} = P_{\alpha\beta}(\delta \to -\delta).
\end{equation}
All the results presented in this section are also valid for
antineutrinos if one makes substitutions
\begin{equation}
    \label{eq:pranti}
    \delta \to -\delta, \quad A_{ij} \to \bar{A}_{ij}, 
    \quad\text{where}\quad
    \bar{A}_{ij} \equiv A_{ij}(V \to -V).
\end{equation}
Notice that the amplitudes of transitions \eqref{eq:Pmue} and
\eqref{eq:Ptaue} that involve $\nu_e$ are given by linear combinations
of two propagation-basis amplitudes. The other amplitudes depend on
three propagation-basis amplitudes.


\subsection{Eigenvalues}

We will refer to parameters $\theta_{12}$ and $\Dmq_{21}$ as the 1-2
(or solar) sector, and to parameters $\theta_{13}$ and $\Dmq_{31}$ as
the 1-3 (or atmospheric) sector. We use the same form of the mixing
matrix in matter as in vacuum, with substitution $\theta_{ij} \to
\theta_{ij}^m$. The eigenvalues of the Hamiltonian $H_{i}^m$ are
identified in such a way that $H_{i}^m \to \Dmq_{i1} / 2E_\nu$ when $V
\to 0$. For densities (energies) between the 1-2 and 1-3 MSW
resonances we find for the normal mass hierarchy
\begin{equation}
    \label{eq:between}
    H_{1}^m \approx \frac{\Dmq_{21} c_{12}^2}{2E_\nu} \,, \qquad
    H_{2}^m \approx V \,, \qquad
    H_{3}^m \approx \frac{\Dmq_{31}}{2E_\nu} \,.
\end{equation}
For high densities (energies) which are far above the 1-3 resonance we
have
\begin{equation}
    \label{eq:above}
    H_{1}^m \approx \frac{\Dmq_{21} c_{12}^2}{2E_\nu} \,, \qquad
    H_{2}^m \approx \frac{\Dmq_{31} c_{13}^2}{2E_\nu} \,, \qquad
    H_{3}^m \approx V \,.
\end{equation}
In a constant density medium the oscillation phases equal
\begin{equation}
    2 \phi_{ji}^m = \Delta H_{ji} \, L
    \qquad\text{with}\qquad
    \Delta H_{ji} \equiv H_{j}^m - H_{i}^m.
\end{equation}
There are two independent frequencies in the problem, $\Delta H_{21}$
and $\Delta H_{32}$.

For antineutrinos there are no level crossings, and with the increase
of density (energy) the eigenvalues have the following asymptotic
limits:
\begin{equation}
    \label{eq:anlevels}
    H_{1}^m \to V \,, \qquad
    H_{2}^m \to \frac{\Dmq_{21} c_{12}^2}{2E_\nu} \,, \qquad
    H_{3}^m \to \frac{\Dmq_{31} c_{13}^2}{2E_\nu} \,.
\end{equation}
We will discuss the level crossing scheme for the inverted hierarchy
in Sec.~\ref{sec:inverted}.


\subsection{Amplitudes in medium of constant density}

Oscillations in a matter of constant (but trajectory dependent)
density layers give a good approximation to the results of exact
numerical calculations for neutrino oscillations in the Earth. As we
have shown in~\cite{Akhmedov:2006hb} they reproduce rather accurately
all the features of the oscillograms (at least for $\Dmq_{21} = 0$).
Thus, we can use the exact analytic results for constant-density
matter to clarify various features of the numerical results for the
realistic Earth density profile. This also allows one to obtain a
parametric dependence of the probabilities, in particular parametric
smallness of certain contributions to the probabilities. In what
follows we present the results for one layer of constant density. We
mark the corresponding amplitudes and probabilities with the
superscript ``$\cst$''.

The exact formula for the $\nu_\mu \to \nu_e$ transition amplitude in
matter of constant density is
\begin{equation}
    \label{eq:me-ampl}
    S_{e \mu}^\cst = 2 i \, e^{i \phi_{21}^m}
    \LT[U_{e1}^{m} U_{\mu 1}^{m *} \sin\phi_{21}^m 
    - e^{-i \phi_{31}^m} U_{e3}^{m} U_{\mu 3}^{m *} \sin\phi_{32}^m
    \RT] \,,
\end{equation}
where $U_{\alpha j}^m$ and $\phi_{ji}^m$ are the elements of mixing
matrix and the oscillation half-phases in matter.
Using the expressions for $U_{e i}^{m}$ and $U_{\mu i}^m$ in terms of
the mixing angles in the standard parametrization, we can rewrite
Eq.~\eqref{eq:me-ampl} as
\begin{equation}
    S_{e \mu}^\cst = \cos\theta_{23}^m A_{e\tilde{2}}^\cst
    + \sin\theta_{23}^m e^{- i\delta^m} A_{e\tilde{3}}^\cst \,,
\end{equation}
where
\begin{align}
    \label{eq:ample2}
    A_{e\tilde{2}}^\cst
    &\equiv -i\, e^{i \phi_{21}^m} \cos\theta_{13}^m \,
    \sin 2\theta_{12}^m \, \sin\phi_{21}^m \,,
    \\
    \label{eq:ample3}
    A_{e\tilde{3}}^\cst
    &\equiv -i \, e^{i \phi_{21}^m} \sin 2\theta_{13}^m
    \LT[\sin \phi_{32}^m \, e^{-i\phi_{31}^m}
    + \cos^2 \theta_{12}^m \sin\phi_{21}^m \RT] \,.
\end{align}
Here $\phi_{31}^m = \phi_{32}^m + \phi_{21}^m$. Since to a good
approximation $\theta_{23}^m \approx \theta_{23}$ and $\delta^m
\approx \delta$~\cite{Toshev:1991ku, Freund:2001pn}, the amplitudes
$A_{e\tilde{2}}^\cst$ and $A_{e\tilde{3}}^\cst$ can be identified with
$A_{e\tilde{2}}$ and $A_{e\tilde{3}}$ in Eq.~\eqref{eq:Pmue} and
\eqref{eq:Ptaue}. According to Eq.~\eqref{eq:between}, between the two
MSW resonances $\phi_{31}^m \approx \Dmq_{32} L / 4 E_\nu$. Above the
1-2 resonance $\sin^2 \theta_{12}^m \approx 1$, so that the second
term in \eqref{eq:ample3} is very small:
\begin{equation}
    \cos^2 \theta_{12}^m \approx \frac{\sin^2 2\theta_{12}}{4}
    \LT(\frac{\Dmq_{21}}{2\, V E_\nu}\RT)^2 =
    \frac{1}{4} \tan^2 2\theta_{12} \LT(\frac{E_{12}^R}{E_\nu}\RT)^2,
\end{equation}
which is suppressed as $1/V^2$. Here $E_{12}^R \approx \cos
2\theta_{12} \Dmq_{21} / 2V$ is the 1-2 resonance energy.
Consequently, up to the phase factor, the amplitude $A_{e\tilde{3}}$
is reduced to the $2\nu$ form with the parameters $(\theta_{13}^m,
\phi_{32}^m)$. Notice that above the 1-3 resonance $\phi_{21}^m \to
\Dmq_{32} L/4E_\nu$.

Similarly, for the $\nu_\mu \to \nu_\mu$ amplitude we obtain
\begin{equation}
    \label{eq:mm-ampl}
    S_{\mu\mu}^\cst = 1
    + 2i\, e^{i \phi_{21}^m} |U_{\mu 1}^m|^2 \sin\phi_{21}^m
    - 2i\, e^{-i \phi_{32}^m} |U_{\mu 3}^m|^2 \sin\phi_{32}^m \,.
\end{equation}
In terms of mixing angles,
\begin{equation}
    \label{eq:represent}
    U_{\mu 1}^m = -s_{12}^m c_{23}^m
    - c_{12}^m s_{13}^m s_{23}^m e^{i\delta^m} \,,
    \qquad U_{\mu 3}^m = c_{13}^m s_{23}^m \,,
\end{equation}
and the amplitude can be rewritten as
\begin{equation}
    \label{eq:mm-ampl-a}
    S_{\mu\mu}^\cst = \cos^2 \theta_{23}^m A_{\tilde{2}\tilde{2}}^\cst
    + \sin^2 \theta_{23}^m A_{\tilde{3}\tilde{3}}^\cst
    + \sin 2\theta_{23}^m \cos\delta^m A_{\tilde{2}\tilde{3}}^\cst \,,
\end{equation}
where
\begin{align}
    \label{eq:ample22}
    A_{\tilde{2}\tilde{2}}^\cst
    &\equiv 1 + 2i\, e^{i \phi_{21}^m} \sin^2 \theta_{12}^m \sin\phi_{21}^m \,,
    \\
    \label{eq:ample33}
    A_{\tilde{3}\tilde{3}}^\cst
    &\equiv 1 - 2i\, e^{-i \phi_{32}^m} \cos^2 \theta_{13}^m \sin\phi_{32}^m
    + 2i\, e^{i \phi_{21}^m} \sin^2\theta_{13}^m
    \cos^2\theta_{12}^m \sin\phi_{21}^m \,,
    \\
    \label{eq:ample23}
    A_{\tilde{2}\tilde{3}}^\cst
    &\equiv i\, e^{i \phi_{21}^m} \sin\theta_{13}^m
    \sin 2\theta_{12}^m \sin\phi_{21}^m \,.
\end{align}
Notice that $A_{\tilde{2}\tilde{2}}^\cst$ has exactly the form of the
corresponding $2\nu$ amplitude driven by the solar parameters. The
amplitude $A_{\tilde{3}\tilde{3}}^\cst$ also coincides to a very good
approximation with the corresponding $2\nu$ amplitude driven by the
atmospheric parameters. The contribution of the solar mode to
$A_{\tilde{3}\tilde{3}}^\cst$ is strongly suppressed by the factor
$\sin^2 \theta_{13}$ in the region of 1-2 resonance and by $\cos^2
\theta_{12}^m$ above this resonance. Up to the overall factor $\sin
\theta_{13}^m$ the amplitude $A_{\tilde{2}\tilde{3}}^\cst$ depends on
the solar (1-2) phase only, and in general it contains double
smallness: $\sin\theta_{13}$ and the one related to the solar mode of
oscillations. Again in the approximation $\theta_{23}^m \approx
\theta_{23}$ and $\delta^m \approx \delta$ the amplitudes
\eqref{eq:ample22}, \eqref{eq:ample33} and \eqref{eq:ample23} can be
identified with the corresponding amplitudes in the propagation basis.

For completeness, we present also the expression for the $\nu_e \to
\nu_e$ amplitude:
\begin{equation}
    \label{eq:ampleee}
    S_{ee}^\cst = 1 + 2i\, e^{i \phi_{21}^m} \cos^2 \theta_{13}^m
    \cos^2 \theta_{12}^m \sin \phi_{21}^m
    - 2i\, e^{-i \phi_{32}^m} \sin^2 \theta_{13}^m \sin\phi_{32}^m.
\end{equation}
It can be rewritten in the form convenient for use at low energies:
\begin{equation}
    \label{eq:ampleee1}
    S_{ee}^\cst = 1 + 2i\, e^{i \phi_{21}^m}
    \cos^2 \theta_{12}^m \sin\phi_{21}^m
    - 2i\, e^{i \phi_{21}^m} \sin^2 \theta_{13}^m
    \LT[ e^{-i \phi_{32}^m} \sin \phi_{31}^m
    + \cos^2 \theta_{12}^m \sin \phi_{21}^m \RT].
\end{equation}

In formulas for the amplitudes one can interchange the phases in the
exponents and sines using the following identity:
\begin{equation}
    \label{eq:phaseexc}
    \sin \phi_{32}^m e^{-i \phi_{31}^m} =
    \sin \phi_{31}^m e^{-i \phi_{32}^m} - \sin \phi_{21}^m,
\end{equation}
where $\phi_{31}^m = \phi_{32}^m + \phi_{21}^m$.  Due to the level
crossing phenomenon this phase interchange is convenient for studies
of oscillation effects in different energy ranges.


\subsection{The factorization approximation}

The elements of the evolution matrix in the propagation basis
$\tilde{S}$ depend in general on $\Dmq_{21}$, $\theta_{12}$,
$\Dmq_{31}$ and $\theta_{13}$. As follows immediately from the form of
the Hamiltonian $\tilde{H}$ in Eq.~\eqref{eq:matr1}, in the limits
$\Dmq_{21} \to 0$ or/and $s_{12} \to 0$ the state $\tilde{\nu}_2$
decouples from the rest of the system, and consequently, the amplitude
$A_{e\tilde{2}}$ vanishes. In this limit, $A_{e\tilde{3}}$ (as well as
$A_{\tilde{3}\tilde{3}}$ and $S_{ee}$) is reduced to a $2\nu$
amplitude which depends on the parameters $\Dmq_{31}$ and
$\theta_{13}$. We denote the latter by $A_A$:
\begin{equation}
    \label{eq:ampl-aa}
    A_A (\Dmq_{31}, \theta_{13}) \equiv
    A_{e\tilde{3}} (\Dmq_{21} = 0) \,.
\end{equation}
It is this amplitude that has been studied in our previous
paper~\cite{Akhmedov:2006hb}; the corresponding probability equals
$P_A = |A_A|^2$.

In the limit $s_{13}\to 0$ the state $\tilde{\nu}_3$ decouples and the
amplitude $A_{e\tilde{3}}$ vanishes. At the same time, the amplitude
$A_{e\tilde{2}}$ reduces to a $2\nu$ amplitude depending on the
parameters of the 1-2 sector, $\Dmq_{21}$ and $\theta_{12}$. Denoting
this amplitude by $A_S$ we have
\begin{equation}
    \label{eq:ampl-as}
    A_S (\Dmq_{21}, \theta_{12})
    \equiv A_{e\tilde{2}}(\theta_{13} = 0) \,.
\end{equation}
We will also use the notation $P_S \equiv |A_S|^2$.

This consideration implies that to the leading non-trivial order in
the small parameters $s_{13}$ and $r_\Delta$ the amplitudes
$A_{e\tilde{2}}$ and $A_{\tilde{2}e}$ below the 1-3 resonance depend
only on the ``solar'' parameters, whereas the amplitudes
$A_{e\tilde{3}}$ and $A_{\tilde{3}e}$ above the 1-2 resonance depend
only on the ``atmospheric'' parameters, \ie:
\begin{equation}\begin{aligned}
    \label{eq:factor}
    A_{e\tilde{2}} \simeq A_{\tilde{2}e}
    &\simeq A_S(\Dmq_{21}, \theta_{12}) \,,
    & \qquad E_\nu &< E_{13}^R \,,
    \\
    A_{e\tilde{3}} \simeq A_{\tilde{3}e}
    & \simeq A_A(\Dmq_{31}, \theta_{13}) \,,
    & \qquad E_\nu &> E_{12}^R \,.
\end{aligned}\end{equation}
In what follows we will call the approximate equalities in 
Eq.~\eqref{eq:factor} the factorization approximation, since the 
dependence of the interference terms in the probabilities, 
$A_{\tilde{2}e}^*A_{\tilde{3}e}$, on solar and atmospheric parameters 
factorizes in this approximation.

An additional insight into the factorization approximation can be
obtained from the results for constant-density matter obtained in the
previous section. According to Eqs.~\eqref{eq:ample2} and
\eqref{eq:ample3}, up to the phase factors one has
\begin{equation}
    \label{eq:2nu-ampl}
    A_A \to A_A^\cst \equiv \sin 2\theta_{13}^m \sin\phi_A, \qquad
    A_S \to A_S^\cst \equiv \sin 2\theta_{12}^m \sin\phi_S \,,
\end{equation}
where
\begin{align}
    \label{eq:o12}
    \phi_S &= \frac{\Dmq_{21} L}{4E_\nu}
    \sqrt{(\cos 2\theta_{12} \mp 2 V E_\nu/\Dmq_{21})^2
      + \sin^2 2\theta_{12}} \,,
    \\
    \label{eq:o13}
    \phi_A &= \frac{\Dmq_{31} L}{4E_\nu}
    \sqrt{(\cos 2\theta_{13} \mp 2 V E_\nu/\Dmq_{31})^2
      + \sin^2 2\theta_{13}} \,.
\end{align}
Here the upper (lower) sign corresponds to neutrinos (antineutrinos).

Due to the level crossing phenomenon the factorization approximation
\eqref{eq:factor} is not valid in the $(E_\nu, \Theta_\nu)$ parameter
space of the 1-3 resonance where 1-3 mixing is enhanced. Indeed, using
Eqs.~\eqref{eq:between} and \eqref{eq:above} we find that for the
normal mass hierarchy
\begin{equation}
    \phi_{32}^m \approx \phi_A \quad\text{for}\quad
    E_\nu \gg E_{12}^R
\end{equation}
and
\begin{equation}
    \phi_{21}^m \approx \phi_S \quad\text{for}\quad
    E_\nu \ll E_{13}^R.
\end{equation}
However, the last formula is not correct in the energy region near the
1-3 resonance and above it due to the 1-3 level crossing. In
particular, from \eqref{eq:above} we obtain for $E_\nu \gg E_{13}^R$
\begin{equation}
    \label{eq:phi21above}
    \phi_{21}^m \approx \frac{L}{4E_\nu}
    \LT( \Dmq_{31} c_{13}^2 - \Dmq_{21} c_{12}^2 \RT)
    \approx \frac{\Dmq_{31} L}{4E_\nu} \equiv \phi_A^0 \,,
\end{equation}
where $\phi_A^0$ is the 1-3 vacuum phase. At the same time, $\phi_S
\approx \phi_{31}^m$.

The results \eqref{eq:ample2} and \eqref{eq:ample3} allow us also to
evaluate corrections to the factorization approximation. As follows
from Eq.~\eqref{eq:ample2}, beyond this approximation the amplitude 
$A_{e\tilde{2}}$ acquires an extra factor $c_{13}^m$ and still depends
on just one oscillation frequency, determined by the solar mass
splitting $\Dmq_{21}$. The factor $\cos\theta_{13}^m$ decreases with
increasing neutrino energy, approaching the value $1/\sqrt{2}$ at the
1-3 MSW resonance and further decreasing to $\cos\theta_{13}^m\simeq
0$ above this resonance. This leads to an additional suppression of
the ``solar'' amplitude $A_{e\tilde{2}}$ in matter at high energies,
on top of the usual suppression due to the quenching of the mixing in
the 1-2 sector.

According to Eq.~\eqref{eq:ample3} the exact expression for
$A_{e\tilde{3}}^\cst$ differs from $A_A^\cst$ by an additional term
depending on the 1-2 frequency. Above the 1-2 resonance we have
$\cos^2 \theta_{12}^m \ll 1$, and this additional term can be omitted:
\begin{equation}
    \label{eq:ample3a}
    A_{e\tilde{3}}^\cst \approx -i\, e^{i \phi_{21}^m}
    \LT[\sin 2\theta_{13}^m \sin\phi_{32}^m
    e^{-i \phi_{31}^m} \RT].
\end{equation}
Hence, up to a phase factor the amplitude $A_{e\tilde{3}}$ is also
reduced to the standard two-neutrino form and depends on a single
oscillation phase $\phi_{32}^m$. We can rewrite the amplitudes
\eqref{eq:ample2} and \eqref{eq:ample3} as
\begin{align}
    \label{eq:ample2a}
    A_{e\tilde{2}}^\cst
    &= -i\, e^{i \phi_{21}^m} \cos\theta_{13}^m A_S^\cst \,
    & (E_\nu &\ll E_{13}^R) \,,
    \\
    \label{eq:ample-con}
    A_{e\tilde{3}}^\cst
    & = -i\, e^{i \phi_{21}^m}\LT[  e^{-i \phi_{31}^m} A_A^\cst +
    \cot^2 \theta_{12}^m \, \sin 2\theta_{13}^m \, A_S^\cst \RT] \,,
    & (E_{12}^R \ll E_\nu &\ll E_{13}^R) \,.
\end{align}

In the general case of a matter of an arbitrary density profile, one
can show, using simple power counting arguments, that the corrections
to the factorization approximation for the amplitude $A_{e\tilde{2}}$
are of order $s_{13}^2$, whereas the corrections to the
``atmospheric'' amplitude $A_{e\tilde{3}}$ are of order
$r_\Delta$~\cite{Takamura:2005df}, in agreement with our consideration
for constant density. The amplitude $A_{e\tilde{3}}$ does not in
general have a 2-flavor form, once the corrections to the
factorization approximation are taken into account.


\section{Effects of 1-2 splitting and mixing}
\label{sec:12split}


\subsection{Neutrino oscillograms of the Earth}

In this section we study the neutrino oscillograms of the Earth
--~contours of constant oscillation probabilities in the plane of
neutrino nadir angle and energy~-- in the complete $3\nu$ context (see
Fig.~\ref{fig:solar}). In our numerical calculations we take the
matter density distribution inside the Earth as given by the PREM
model~\cite{Dziewonski:1981xy}. Unless otherwise specified, we assume
the normal neutrino mass hierarchy and use the following numerical
values of neutrino parameters: $\Dmq_{31} = 2.5 \cdot 10^{-3}~\eVq$,
$\Dmq_{21} = 8 \cdot 10^{-5}~\eVq$, $\tan \theta_{12} =
0.45$~\cite{GonzalezGarcia:2007ib, Maltoni:2004ei, Fogli:2005cq,
Strumia:2006db}.

Recall that the distance $L$ that neutrinos propagate inside the Earth
is related to the nadir angle of the neutrino trajectory $\Theta_\nu$
through
\begin{equation}
    L = 2 R_\oplus \cos\Theta_\nu \,,
\end{equation}
where $R_\oplus = 6371$~km is the Earth radius. The value $\Theta_\nu
= 0$ corresponds to vertically up-going neutrinos which travel along
the Earth's diameter, whereas $\Theta_\nu = \pi/2$ corresponds to
horizontal neutrino trajectories. For $0\le \Theta_\nu \le 33^\circ$
neutrinos cross both the mantle and the core of the Earth, whereas for
larger values of the nadir angle they only cross the Earth's mantle.
We call the corresponding parts of the oscillograms the core domain
and the mantle domain, respectively.

\PAGEFIGURE{
  \includegraphics[width=141mm]{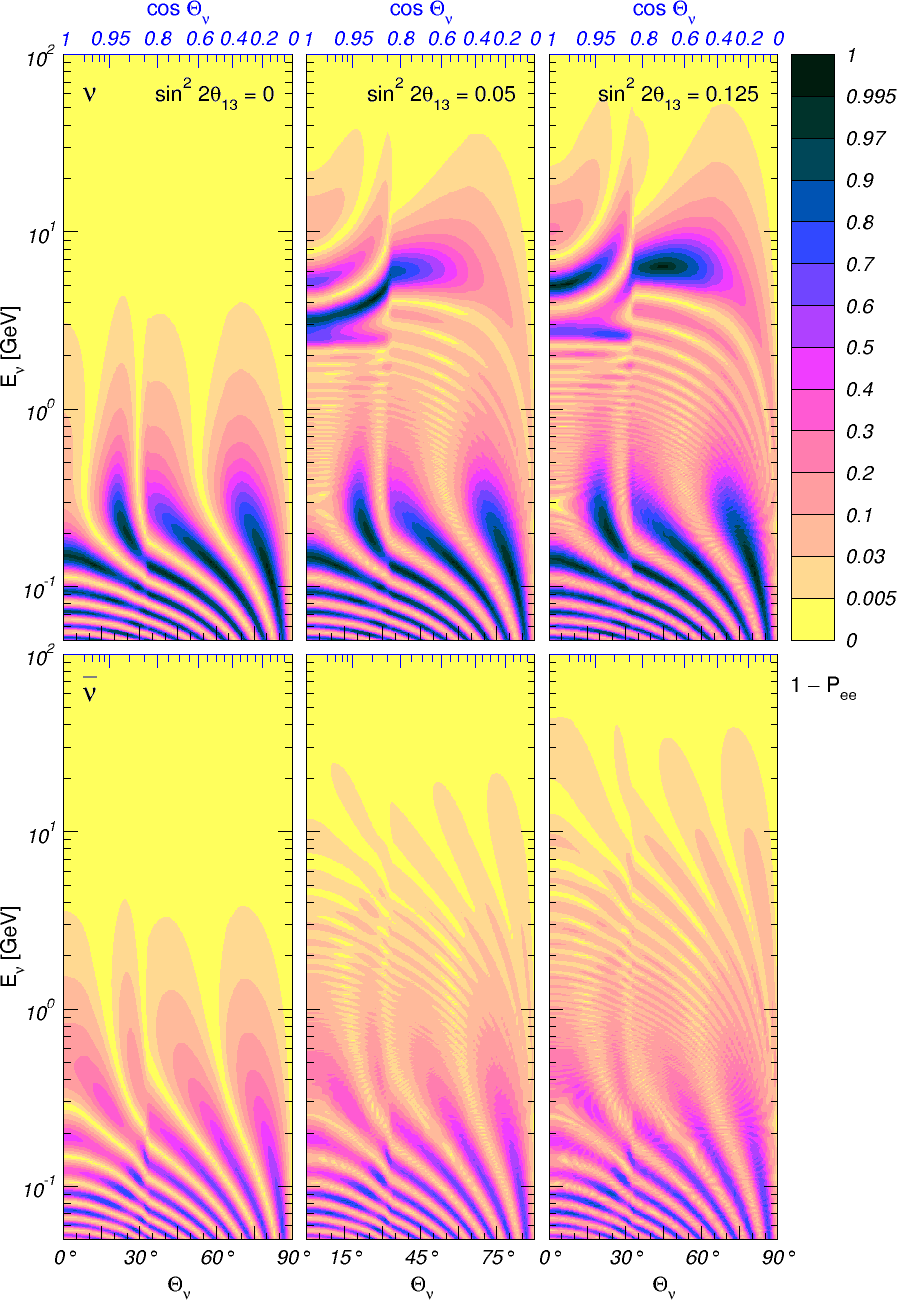}
  \caption{\label{fig:solar}%
    Neutrino oscillograms in the $3\nu$-mixing case. Shown are the
    contours of constant probability $1 - P_{ee}$ (upper panels) and
    $1 - P_{\bar{e}\bar{e}}$ (lower panels) for $\Dmq_{21} = 8 \times
    10^{-5}~\eVq$, $\tan^2\theta_{12} = 0.45$ and different values of
    $\theta_{13}$.}}

In the limit of vanishing $\Dmq_{21}$ the main features of the
neutrino oscillations in the Earth are determined by the MSW resonance
enhancement of neutrino oscillations and the parametric resonance
enhancement in the core domain (see Ref.~\cite{Akhmedov:2006hb} and
Fig.~\ref{fig:solar}). The former leads to the appearance of the MSW
resonance peaks: one in the mantle domain of the oscillogram at
neutrino energies $E_\nu \sim (6-7)~\GeV$, and another one in the core
domain at $E_\nu\sim (2-3)~\GeV$. The latter (parametric enhancement)
produces three parametric ridges in the core domain.  In the notation
of Ref.~\cite{Akhmedov:2006hb} they are ridges A, B, C. Ridge A spans
the energy range from $E_\nu \approx 3~\GeV$ at $\Theta_\nu = 0$ to
$E_\nu \approx 6~\GeV$ at the mantle-core border, where it merges with
the mantle MSW resonance peak (see Fig.~\ref{fig:solar} middle and
right panels). Ridge B starts from $E_\nu \approx 5~\GeV$ at
$\Theta_\nu = 0$ and becomes nearly vertical for $\Theta_\nu =
33^\circ$. The energy of the ridge C increases from $E_\nu \approx
10~\GeV$ to high energies when the nadir angle changes from
$\Theta_\nu = 0$ to $\Theta_\nu = 33^\circ$. The location of the
ridges weakly depends on the 1-3 mixing angle. The ridges differ by
the oscillation phase acquired in the core~\cite{Akhmedov:2006hb}.


\subsection{Oscillograms due to 1-2 mixing}

The oscillograms for $\theta_{13} = 0$ are presented in the left
panels of Fig.~\ref{fig:solar}.  In this limit, according to
Eqs.~\eqref{eq:matr3} and \eqref{eq:ampl-as},
\begin{equation}
    1 - P_{ee} = |A_{e\tilde{2}}|^2 = P_S.
\end{equation}
Thus, shown are the contours of constant probability $P_S$.

In the $2\nu$ case the oscillation probabilities depend on $\Dmq$ and
neutrino energy $E_\nu$ via the ratio $\Dmq/E_\nu$. Therefore when
oscillations are driven by the solar splitting, $\Dmq_{21}$, the
oscillation pattern is shifted to smaller energies as compared to that
due to $\Dmq_{31}$. Moreover, the 1-2 pattern differs from the pattern
for the 1-3 mixing due to the large value of the 1-2 mixing. Indeed,
as a consequence of the large 1-2 mixing the following new features
appear.
\begin{myitemize}
  \item[1.] The oscillation length at the resonance is smaller than
    that for small mixing
    \begin{equation}
	\label{eq:lmres}
	l_m^R = \frac{l_\nu}{\sin 2\theta_{12}} \sim l_\nu \,,
    \end{equation}
    where $l_\nu$ is the vacuum oscillation length.
    
  \item[2.] The resonance energy is shifted to smaller values not only
    due to $\Dmq_{21} \ll \Dmq_{31}$ but also because of the factor
    $\cos 2\theta_{12} \approx 0.4$:
    \begin{equation}
	\label{eq:eres12}
	E_{12}^R = \frac{\Dmq_{21}}{2 \bar{V}} \cos 2\theta_{12} \,.
    \end{equation}
    Here $\bar{V}$ is an effective (average) value of potential.
    
  \item[3.] The degree of adiabaticity is better than for the 1-3
    mixing case, and therefore the oscillation probability in the
    mantle is determined by the potential near the surface of the
    Earth, $\bar{V}$, averaged over the distance of the order of
    oscillation length.
    
  \item[4.] The oscillation length in matter, $l_m = 2\pi / \Delta
    H_{21}$, monotonically increases with energy, approaching in the
    limit $E_\nu \to \infty$ the refraction length $l_0 \equiv
    2\pi/V$. Recall that for small mixings $l_m$ first increases with
    energy, reaches a maximum slightly above the resonance and then
    decreases.
    
  \item[5.] The jump of the mixing angle at the mantle-core boundary is
    small. Therefore, a sudden distortion of the oscillation patterns
    at $\Theta_\nu = 33^\circ$ is not as significant as it is for the
    small 1-3 mixing, especially below the 1-2 resonance energy.
\end{myitemize}
These features allow one to understand the structure of the
oscillograms. In the mantle domain ($\Theta_\nu > 33^\circ$) the
oscillation pattern for neutrinos is determined by the resonance
enhancement of oscillations. There are three MSW resonance peaks above
$0.1~\GeV$, which differ from each other by the value of the total
oscillation phase. The outer peak ($\Theta_\nu \approx 82^\circ$)
corresponds to $\phi \approx \pi/2$, the deeper one at $\Theta_\nu =
60^\circ$, to $\phi \approx 3\pi/2$, and the inner one ($\Theta_\nu
\approx 40^\circ$), to $\phi = 5\pi/2$. Recall that such a large phase
can be acquired due to a smaller resonance oscillation length
\eqref{eq:lmres} in comparison to the length in the 1-3 mixing case,
in which only one peak with $\phi = \pi/2$ can be realized (see the
upper parts of the panels in Fig.~\ref{fig:solar}). The resonance
energy is given by Eq.~\eqref{eq:eres12}, and for the surface
potential we find
\begin{equation}
    \label{eq:eressurf}
    E_{12}^R \approx 0.12~\GeV \,.
\end{equation}
The ratio of the 1-2 and 1-3 resonance energies equals
\begin{equation}
    \frac{E_{12}^R}{E_{13}^R} =
    \LT( \frac{\Dmq_{21}}{\Dmq_{31}} \RT)
    \LT( \frac{\cos 2\theta_{12}}{\cos 2\theta_{13}} \RT)
    \LT( \frac{\bar{V}_{13}}{\bar{V}_{12}} \RT)
    \approx \frac{1}{50} \,.
\end{equation}
Here $\bar{V}_{13} / \bar{V}_{12} \approx 1.5$
(see~\cite{Akhmedov:2006hb}), since for 1-3 oscillations we should
take the average of potential along the whole trajectory. The estimate
\eqref{eq:eressurf} is valid for two outer peaks. For the peak at
$\Theta_\nu = 40^\circ$, $\bar V$ is larger, and accordingly, the
resonance energy is slightly smaller.

The width of the 1-2 resonance is larger and therefore the regions of
sizable oscillation probability are more extended in the $E_\nu$
direction as compared to the oscillations governed by the 1-3 mixing
and splitting.

\FIGURE[!t]{
  \includegraphics[width=142mm]{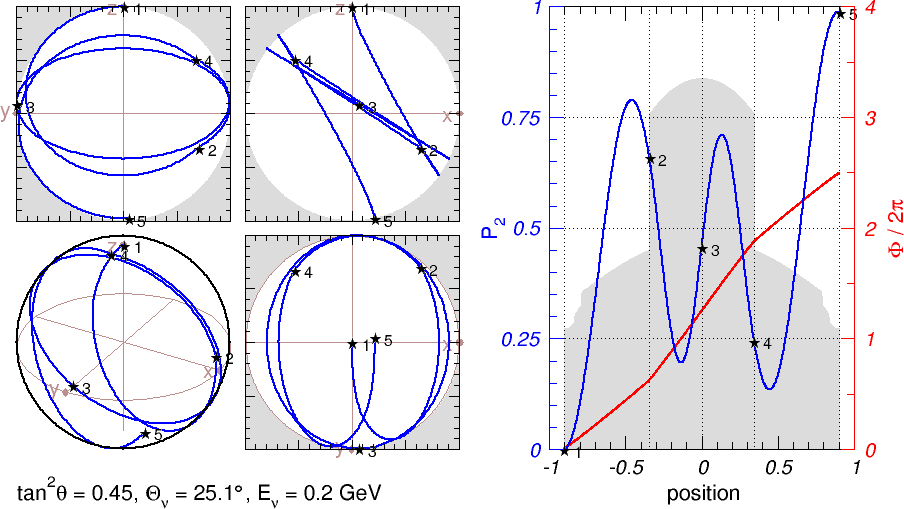}
  \caption{\label{fig:evolution}%
    Parametric resonance due to 1-2 mixing and splitting at the
    parametric peak at $\Theta_\nu = 25.1^\circ$ and $E_\nu =
    0.2~\GeV$. Left panels: trajectory of the neutrino polarization
    vector in the flavor space, and projections of this trajectory
    onto three planes. Right panel: dependence of $P_S$ (blue line)
    and of the oscillation phase (red line) on the distance along the
    neutrino trajectory. The stars along the blue line correspond to
    the beginning (1), the mantle-core boundary (2), the midpoint (3),
    the core-mantle boundary (4), and the end of the trajectory (5).}}

The resonance energy in the core is $E_{12}^R \approx 0.04~\GeV$.
Therefore for $E_\nu > (0.10-0.15)~\GeV$ the 1-2 mixing in the core is
substantially suppressed by matter. Furthermore, at the energies above
the resonance energy in the mantle ($E_\nu > 0.12~\GeV$) the mixing in
the mantle is also suppressed. Therefore the peaks with $P_{max}
\approx 1$ at $E_\nu > 0.12~\GeV$ should be due to the interplay of
the core and mantle effects. In particular, the peak at $E_\nu \simeq
0.2~\GeV$ and $\Theta_\nu \simeq 25^\circ$ is due to the parametric
enhancement of the oscillations. It corresponds to the realization of
the parametric resonance condition when the oscillation half-phases
equal approximately $\phi_\text{mantle} \approx \pi/2$ and
$\phi_\text{core} \approx 3\pi/2$ (note that the total phase $\approx
5\pi/2$, and this parametric ridge is attached to the $5\pi/2$ - MSW
peak in the mantle domain). The spatial evolution of neutrinos in this
peak and its graphical representation are shown in
Fig.~\ref{fig:evolution}. In the left panel the blue lines present the
trajectory of the neutrino polarization vector in the flavor space.
Recall that in terms of the elements of the $2 \times 2$ evolution $S$
matrix, $S_{11}$, $S_{12}$, $S_{21} = -S_{12}^* $ and $S_{22} =
S_{11}^*$ (where the last two equalities are the consequences of
unitarity and hold in the basis where the effective Hamiltonian is 
traceless), the polarization vector is defined in the flavor space as
the vector with components $s_X = \Re[S_{11}^* S_{12}]$, $s_Y =
\Im[S_{11}^* S_{12}]$, $s_Z = |S_{11}|^2 -1/2$. Then the $\nu_e$
survival probability is given by $P_{ee} = |S_{11}|^2 = s_Z + 1/2$
(see~\cite{Akhmedov:2006hb} for details).

\FIGURE[!t]{
  \includegraphics[width=143mm]{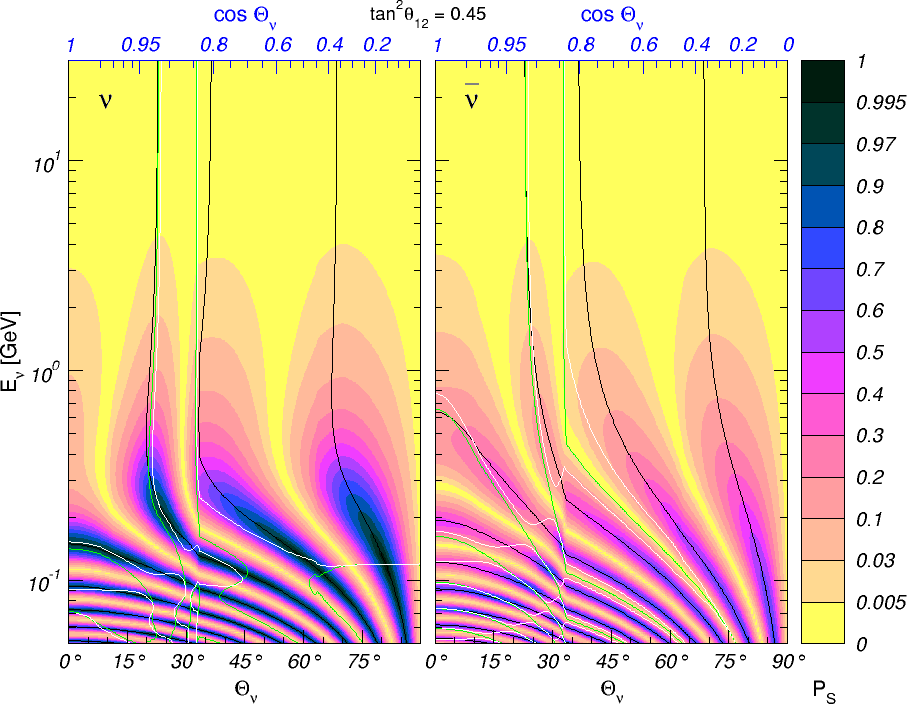}
  \caption{\label{fig:collinear}%
    Interpretation of the oscillograms due to the 1-2 mixing and mass
    splitting, for neutrinos (left) and antineutrinos (right). Shown
    (in color) are contours of constant $P_S$ as well as lines of
    various conditions that explain different structures of the
    oscillograms. The lines correspond to collinearity condition
    (white), the generalized resonance condition (green), and the
    phase condition (black).}}

The ridge at $E_\nu \simeq (0.12-0.15)~\GeV$ and $\Theta_\nu \simeq
0-12^\circ$ can also be considered as being due to the parametric
resonance with a larger core phase: $\phi_\text{core} \approx 5\pi/2$.
However, for energies $E_\nu \leq 0.15~\GeV$ the in-matter mixing is
nearly maximal both in the mantle and core, and the effect of the
mantle-core density jump is small. As follows from
Fig.~\ref{fig:collinear} (left panel), the positions of the MSW peaks
and maxima of the parametric ridges are determined very well by the
intersections of the lines that correspond to the collinearity
condition and the phase condition as in the case of oscillograms due
to the 1-3 mixing~\cite{Akhmedov:2006hb}. Recall that in terms of the
elements of the evolution matrices in the mantle, $S^m$, and in the
core, $S^c$, these conditions read 
\begin{equation}
    \Re[S_{11}^m S_{11}^c S_{12}^m] = 0 \text{~(collinearity),}
    \qquad
    \Re[({S^m}^T S^c S^m)_{11}] = 0 \text{~(phase)}.
\end{equation}
In Fig.~\ref{fig:collinear} shown are also the lines of the
generalized resonance condition for symmetric density profile:
$\Im[(S^c S^m)_{11}] = 0$ (see~\cite{Akhmedov:2006hb} for details).
The elements of the evolution matrix have been found by precise
numerical computations.  Apparently the ridges lie along the lines of
the collinearity condition.

There are also several intersections of the collinearity and the phase
condition lines in the core domain of the antineutrino oscillogram
(right panel). This shows the existence of the parametric enhancement
in the antineutrino channel.

For $E_\nu > 0.3~\GeV$ the oscillation length practically does not
depend on the neutrino energy and is close to the refraction length,
$l_0 = 2\pi / (\sqrt{2} G_F N_e)$. Therefore the lines of equal phases
become nearly vertical. According to Fig.~\ref{fig:collinear}, the
lines of zero oscillatory factor, $\phi = \pi k$ with $k = 1, 2, 3$,
that determine the so called ``solar'' magic lines
(see~\cite{Smirnov:2006sm}) are at $\Theta_\nu \sim 54^\circ$ in the
mantle and at $\sim 30^\circ$ and $12^\circ$ in the core domain. The
oscillation probabilities become smaller than $0.5\%$ above $4~\GeV$
in the whole range of the nadir angles.

At lower energies, $E_\nu < 0.1~\GeV$, one finds a regular oscillatory
pattern with ridges and valleys. The distortion of this pattern at the
core-mantle boundary is rather weak due to the smallness of the
difference between the mixing angles in the mantle and core.

In the $2\nu$ context the oscillation probabilities depend on $E_\nu /
\Dmq$ and the mixing angle. Therefore with the increase of the mixing
angle the oscillatory pattern obtained for the 1-3 mixing (upper parts
of the oscillograms) should continuously transform into the pattern
due to the 1-2 mixing (apart from the trivial shift of energy).  We
find that with increasing $\theta$ the parametric ridge A transforms
first into the MSW peak in the mantle (1-3 mixing) and then to the
outer MSW peak of the 1-2 pattern. The parametric ridge B transforms
into the second MSW peak. The 1-3 core peak splits into two parts. One
part transforms into the third MSW peak of the 1-2 pattern in the
mantle. The second part merges with the parametric ridge C and appears
as the parametric peak in the core domain at $E_\nu = 0.2~\GeV$ at
large mixings.

At high energies the patterns for neutrinos and antineutrinos are
rather similar.

Turning on the non-zero 1-3 mixing leads to the appearance of an 1-3
oscillation pattern at high energies and to the interference of the
1-2 and 1-3 oscillation modes. We will discuss two types of the
interference. The first one is the interference of modes characterized
by the solar and atmospheric frequencies. The corresponding
interference terms in probabilities do not necessarily depend on the
CP-violating phase. The second type yields the interference terms
which depends on the CP-violating phase. We will call this the
CP-interference.

In the following subsections we will consider the effects of the
inclusion of the 1-2 mixing onto the 1-3 oscillatory pattern in
different oscillation channels. We will compare the probabilities
computed in the three-flavor ($3\nu$) context and in the two-flavor
($2\nu$) limit $\Dmq_{21} \to 0$, $\theta_{12} \to 0$. The $2\nu$
probabilities are computed as $P(\Dmq_{31}, \sin^2 2\theta_{13})$.
That is, we take a single mass splitting in the $2\nu$ context to
coincide with the largest mass splitting in the $3\nu$ case (the
normal mass hierarchy).  The oscillograms are computed for $\delta =
0$.


\subsection{$\nu_e - \nu_e$ channel}

In Fig.~\ref{fig:solar} (upper panels) we show the probability
$1-P_{ee}$ for three different values of $\sin^2 2\theta_{13}$: zero
(left), small (middle) and relatively large (right). As follows from
the figure, in the first approximation the oscillograms for non-zero
values of $\theta_{13}$ appear as superposition of the $2\nu$
oscillation patterns produced by the 1-2 and 1-3 mixings with small
interference effects.
The smallness of the interference terms for this channel can be
understood in the following way. According to Eq.~\eqref{eq:Pmue},
\eqref{eq:Ptaue} and \eqref{eq:Trev}, the total probability of the
$\nu_e$ disappearance is
\begin{equation}
    \label{eq:3-tran}
    1 - P_{ee} = P_{e\mu} + P_{e\tau}
    = P_{e\tilde{2}} + P_{e\tilde{3}}.
\end{equation}
When the 1-2 splitting is neglected, this probability reduces to
$P_A$, studied in detail in~\cite{Akhmedov:2006hb}.
Eq.~\eqref{eq:3-tran} shows that for $\Dmq_{21} \ne 0$ the $\nu_e -
\nu_{\tilde{2}}$ and $\nu_e - \nu_{\tilde{3}}$ transition
probabilities simply add up in $1 - P_{ee}$ and no interference
between the corresponding amplitudes (no CP-interference) occurs.
Correspondingly, the probability $P_{ee}$ does not depend on the
CP-violating phase in the standard parametrization. It does not depend
on the 2-3 mixing either.

It follows from Eqs.~\eqref{eq:ample2} and \eqref{eq:ample3} that
$P_{e\tilde{2}}$ is driven only by the ``solar'' frequency ($\propto
\Dmq_{21}$), at least in the constant-density approximation, whereas
$P_{e\tilde{3}}$ depends both on the ``atmospheric'' parameters
($\Dmq_{31}$, $\theta_{13}$) and on the parameters of the 1-2 sector.
Therefore the interference of the solar and atmospheric modes
originates from $P_{e\tilde{3}} \equiv |A_{e\tilde{3}}|^2$. The
interference of 1-2 and 1-3 oscillation modes can be quantified (at
least for the mantle region) using the expression \eqref{eq:ample3}
valid for constant density matter:
\begin{equation}
    \label{eq:ee-const}
    1 - P_{ee}^\cst = \cos^2 \theta_{13}^m
    (1 + \sin^2 \theta_{13}^m \, \cot^2 \theta_{12}^m) P_S^\cst
    + P_A^\cst + \cot\theta_{12}^m \, \sin 2\theta_{13}^m \,
    \cos\phi_{31}^m \, A_A^\cst \, A_S^\cst \,.
\end{equation}
Here $A_A^\cst$ and $A_S^\cst$ are the $2\nu$ amplitudes defined in
\eqref{eq:2nu-ampl} (note that in terms of $A_A^\cst$ and $A_S^\cst$
this expression is valid in the energy range between the two
resonances. In the other energy ranges one needs to take into account
the level crossing phenomenon, which changes the labeling of the
phases). The last term in \eqref{eq:ee-const} is due to the
interference of the solar and atmospheric modes which comes from
$|A_{e\tilde{3}}|^2$. Apart from the product $A_A^\cst A_S^\cst$, this
term contains additional small factors. In the region of the 1-2
resonance, $\theta_{13}^m \approx \theta_{13}$, $\cot \theta_{12}^m
\sim 1$ and therefore the interference term is suppressed by a small
factor $\sin 2\theta_{13}$. In fact, all corrections to the main
contribution, $P_S$, are of the order of the small probability
$P_A^\cst \sim \sin^2 \theta_{13}$.
In the region of the 1-3 resonance $P_S^\cst$ is small, and the
interference term is suppressed by $\cot \theta_{12}^m \sim r_\Delta$.
Consequently, all the corrections to the dominant term $P_A^\cst$ are
of the order of $P_S^\cst$. The interference term is further
suppressed at the energies between the two resonances. Indeed, we can
rewrite this term approximately as $\frac{1}{2} \sin 2\theta_{12}^m
\sin 2\theta_{13}^m \cos\phi_{31}^m A_A^\cst A_S^\cst$. As we will see
in Sec.~\ref{sec:sens-me} for small $\theta_{13}$ the product $\sin
2\theta_{12}^m \sin 2\theta_{13}^m$ has a minimum between the two 
resonances. Thus, the strong suppression of the effects of the 1-2
mixing in the $\nu_e - \nu_e$ channel is due to the absence of the
CP-interference of the amplitudes $A_{e\tilde{2}}$ and
$A_{e\tilde{3}}$.

\FIGURE[!t]{
  \includegraphics[width=147mm]{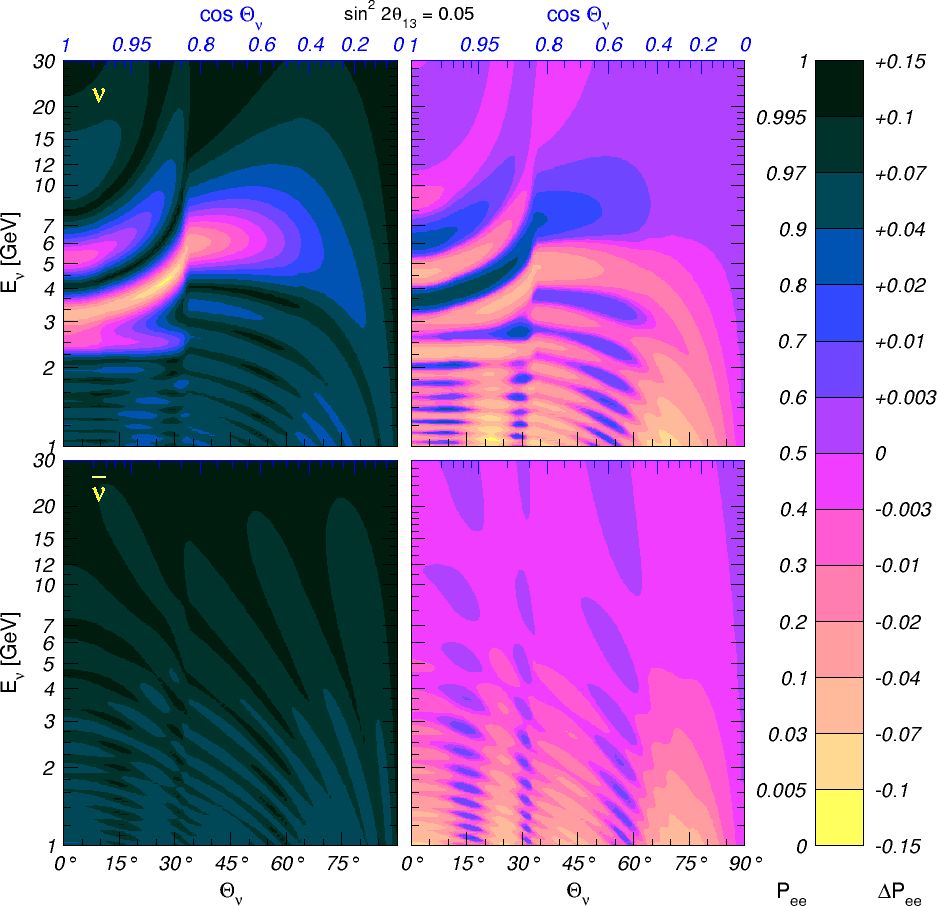}
  \caption{\label{fig:pee-N}%
    Oscillograms for the $\nu_e - \nu_e$ channel. Shown are the
    contours of constant probability $P_{ee}$ (left) as well as
    constant difference $\Delta P_{ee}$ of $3\nu$ and $2\nu$
    probabilities (right), for neutrinos (upper panels) and
    antineutrinos (lower panels). The oscillation parameters for
    $3\nu$ probabilities are $\sin^2 2\theta_{13} = 0.05$, $\Dmq_{21}
    = 8 \times 10^{-5}~\eVq$, $\tan^2\theta_{12} = 0.45$ and $\delta =
    0$. For the $2\nu$ probabilities we used $\Dmq = \Dmq_{31}$.}}

To further illustrate the effects of the 1-2 mixing and mass splitting
at high energies we show in Fig.~\ref{fig:pee-N} the oscillograms for
$P_{ee}$ in the full three-flavor framework (left panels) and for the
difference of probabilities with and without 1-2 mixing and splitting
(right panels):
\begin{equation}
    \Delta P_{ee} \equiv P_{ee} - \mathring{P}_{ee}
    = P_{ee} - P_{ee}(\Dmq_{21} = 0).
\end{equation}
Recall that we compute the two-flavor probability $\mathring{P}_{ee}$
taking $\Dmq = \Dmq_{31}$.

In general, there are two contributions to $\Delta P_{ee}$:
\begin{equation}
    \label{eq:dee2}
    \Delta P_{ee} \approx \Delta P_{ee}^S + \Delta P_{ee}^A.
\end{equation}
which we will refer to as solar and atmospheric contributions. The
solar contribution $\Delta P_{ee}^S$ is proportional to the solar
amplitude, it includes term $P_{ee}^S$ and the interference of the
amplitudes with the solar and atmospheric frequencies in
$P_{e\tilde{3}}$. The second contribution in \eqref{eq:dee2} follows
from a change of the atmospheric mode: the phase and the amplitude of
oscillations due to non-zero 1-2 mass splitting and mixing:
\begin{equation}
    \Delta P_{ee}^A \approx \Delta P^A \equiv P_A - \mathring{P}_A \,.
\end{equation}

Let us estimate these contributions using the results obtained for a
matter of constant density. From Eq.~\eqref{eq:ee-const} we have
\begin{equation}
    \Delta P_{ee}^S \approx \cos^2 \theta_{13}^m \, P_S^\cst +
    \cos \theta_{12}^m \, \sin 2\theta_{13}^m \, \cos\phi_{31}^m \,
    A_A^\cst \, A_S^\cst \,,
\end{equation}
and, according to our consideration above, for high energies $\Delta
P_{ee}^S = \mathcal{O}(P_S^\cst) \leq \sin^2 2\theta_{12}^m$, which is
below $0.002$ in the 1-3 resonance region. For the atmospheric
contribution we obtain
\begin{equation}
    \label{eq:diff}
    \Delta P^A \approx \sin^2 2\mathring{\theta}_{13}^m \,
    \sin^2 \mathring{\phi}_{31}^m
    - \sin^2 2\theta_{13}^m \, \sin^2 \phi_{32}^m \,.
\end{equation}
Let us underline that due to the 1-2 level crossing, in the $3\nu$
case the relevant atmospheric phase is $\phi_{32}^m$ and not
$\phi_{31}^m$. In the lowest order
\begin{equation}
    \label{eq:theta-dfl}
    \sin^2 2\theta_{13}^m
    \approx \sin^2 2\mathring{\theta}_{13}^m(V \to V_1),
\end{equation}
where
\begin{equation}
    \label{eq:vchange}
    V_1 \approx \frac{V}{1 - s_{12}^2 r_\Delta}.
\end{equation}
Then, using Eqs.~\eqref{eq:theta-dfl} and~\eqref{eq:vchange}, we
obtain
\begin{equation}
    \label{eq:dpa}
    \Delta P^A \approx \sin^2 2\mathring{\theta}_{13}^m
    \bigg[ \Delta \phi \sin 2 \phi_{32}^m
    - 2\, \frac{\sin^2 2\mathring{\theta}_{13}^m}{\sin^2 2\theta_{13}}\sin^2 \phi_{32}^m
    \LT( \cos 2\theta_{13} - x \RT) x s_{12}^2 r_\Delta \bigg] \,,
\end{equation}
where
\begin{equation}
    \label{eq:def-x}
    x \equiv \frac{2V E_\nu}{\Dmq_{31}} \,.
\end{equation}
The first term in the brackets is proportional to the phase shift
\begin{equation}
    \Delta \phi = \mathring{\phi}_{31}^m - \phi_{32}^m
    \sim \frac{\Dmq_{21} L}{2E_\nu} \,,
\end{equation}
while the second one is due to the modification of the mixing angle.
The doubly suppressed corrections, $\sim \Delta\phi r_\Delta$, are
omitted. We find that for $\sin^2 2\theta_{13} = 0.05$ both terms are
of order $0.02-0.03$ in the region of the 1-3 resonance (in some
regions of the nadir angles the second term can dominate).

The difference $\Delta P^A$ can be rewritten as
\begin{equation}
    \label{eq:dpa1}
    \Delta P^A \approx 2 A_A^\cst \sin 2\mathring{\theta}_{13}^m
    \bigg[ \cos\phi_{32}^m \, \Delta\phi
    - \frac{\sin^2 2\mathring{\theta}_{13}^m}{\sin^2 2\theta_{13}}
    \sin\phi_{32}^m \,
    \LT(\cos 2\theta_{13} - x \RT)
    x s_{12}^2 r_\Delta \bigg] \,,
\end{equation}
so that its proportionality to $A_A^\cst$ becomes manifest.
Consequently, $\Delta P^A$ vanishes along the atmospheric magic lines,
$A_A^\cst = 0$~\cite{Smirnov:2006sm} (see Sec.~\ref{sec:magic} for
details).

Thus, for the $\nu_e - \nu_e$ channel, $\Delta P_{ee}^A$ dominates
over $\Delta P_{ee}^S$ at high energies and therefore it describes the
structure of the oscillograms for the probability differences. In
particular, this explains the fact that oscillograms for $\Delta
P_{ee}$ repeat the structure of $P_{ee}$ with certain shift in energy.
Partly the difference $\Delta P_{ee}$ can be eliminated by modifying
the $2\nu$ value of $\Dmq_{31}$, taking $\Dmq \neq \Dmq_{31}$. In
certain energy range the phase shift effect can be eliminated. Notice
also that $\Delta P_{ee}^A$ is not proportional to $A^S$ or even to
the corresponding oscillatory factor.

With the decrease of neutrino energy, the effect of the 1-2 mixing
increases. Since $\Delta P^A \propto E_\nu^{-1}$ and $\Delta P^S
\propto E_\nu^{-2}$, at lower energies the effect of the solar
contribution becomes important. The interference effects of different
modes are suppressed in $P_{ee}$, therefore the order of magnitude of
the contributions to $1-P_{ee}$ due to nonzero 1-2 splitting can be
readily estimated from the upper left panel of Fig.~\ref{fig:solar}. 

The oscillograms for the antineutrino channel $\bar{\nu}_e \to
\bar{\nu}_e$ are shown in the lower panels of Figs.~\ref{fig:solar}
and~\ref{fig:pee-N}. Now apart from the change of the sign of $\delta$
one needs to take into account the change of the mixing and of the
level crossing scheme. According to \eqref{eq:anlevels}, the phases
become $\phi_A \simeq \phi_{31}^m$ and $\phi_S \simeq \phi_{21}^m$.
Therefore, the relevant oscillation phase in $A_{e\tilde{3}}^\cst$ is
$\phi_{31}^m$, and the amplitude \eqref{eq:ample3} can be rewritten
for antineutrinos as
\begin{equation}
    \label{eq:ample3anti}
    \bar{A}_{e\tilde{3}}^\cst = ie^{i \phi_{21}^m} \sin 2\theta_{13}^m
    [\sin \phi_{31}^m e^{-i\phi_{32}^m} - \sin^2 \theta_{12}^m
    \sin\phi_{21}^m] \,,
\end{equation}
where we used the phase exchange relation \eqref{eq:phaseexc}. At high
energies $\sin^2 \theta_{12}^m \ll 1$, and the second term in
\eqref{eq:ample3anti} is very small. For the normal mass hierarchy
both 1-2 and 1-3 mixings monotonically decrease with energy. The
effect the of inclusion of the 1-2 mixing and splitting is illustrated
in the lower panels of Fig.~\ref{fig:pee-N}. Since for the
antineutrino channel, in the absence of level crossing, the phase
$\phi_{31}^m$ is relevant, the difference $\mathring{\phi}_{31}^m -
\phi_{31}^m$ is smaller than the corresponding difference in the
neutrino channel. Thus, both $\Delta P_A$ and the total difference of
the probabilities turn out to be smaller than they are in the neutrino
channel. Furthermore, the solar contribution plays an important role
now, determining the vertical structure of the oscillogram.


\subsection{$\nu_e - \nu_\mu$ and $\nu_e - \nu_\tau$ channels}

The transition probability $P_{\mu e} \equiv P(\nu_\mu \to \nu_e)$
(see \eqref{eq:Pmue}) can be rewritten as
\begin{equation}
    \label{eq:Pmue1}
    P_{\mu e} = c_{23}^2 |A_{e\tilde{2}}|^2
    + s_{23}^2 |A_{e\tilde{3}}|^2
    + \sin 2\theta_{23} |A_{e\tilde{2}}^* A_{e\tilde{3}}|
    \cos(\phi - \delta) \,,
\end{equation}
where
\begin{equation}
    \label{eq:arg}
    \phi \equiv \arg(A_{e\tilde{2}}^* A_{e\tilde{3}}) \,.
\end{equation}
Unlike $1-P_{ee}$, this probability contains the interference term
between $A_{e\tilde{2}}$ and $A_{e\tilde{3}}$ which depends on the
CP-violation phase. Furthermore, this interference term is not
suppressed by additional small factors as it happens in the $\nu_e -
\nu_e$ channel. From the unitarity of the matrix $\tilde{S}$ in
Eq.~\eqref{eq:matr2} we obtain for the product of amplitudes in the
interference term
\begin{equation}
    A_{e\tilde{2}}^* A_{e\tilde{3}}
    = - A_{\tilde{3}\tilde{2}}^* A_{\tilde{3}\tilde{3}}
    - A_{\tilde{2}\tilde{3}} A_{\tilde{2}\tilde{2}}^* \,,
\end{equation}
\ie, $A_{e\tilde{2}}^* A_{\tilde{3}e}$ is proportional to the small
amplitudes $A_{\tilde{2}\tilde{3}}$ and $A_{\tilde{3}\tilde2}$.

Since the amplitude $A_{e\tilde{2}}$ is suppressed at high energies
due to the smallness of the 1-2 mixing in matter, in the lowest
approximation we have
\begin{equation}
    P_{\mu e} \approx \sin^2 \theta_{23} |A_{e\tilde{3}}|^2
    \approx \sin^2 \theta_{23} |A_A|^2.
\end{equation}
The maximal value of the probability $P_{\mu e} \simeq s_{23}^2 \simeq
0.5$.

In the constant density approximation (valid for the mantle domain),
using the amplitudes \eqref{eq:ample3} we find at high energies
\begin{equation}
    \label{eq:me-cons}
    P_{\mu e}^\cst = s_{23}^2 P_A^\cst
    + \sin 2 \theta_{23} \cos\theta_{13}^m
    A_A^\cst A_S^\cst \cos (\phi_{31}^m + \delta) + \mathcal{O}(P_S^\cst).
\end{equation}
Comparing this expression with \eqref{eq:Pmue1} we find that for
energies between the two resonances the interference phase satisfies
\begin{equation}
    \label{eq:relphase}
    \phi \approx - \phi_{31}^m \,.
\end{equation}
The precise expression for the phase in the constant density
approximation can be obtained from $\phi \equiv
\arg({A_{e\tilde{2}}^\cst}^* A_{e\tilde{3}}^\cst) =
\arg(A_{e\tilde{3}}^\cst)$.

\FIGURE[!t]{
  \includegraphics[width=147mm]{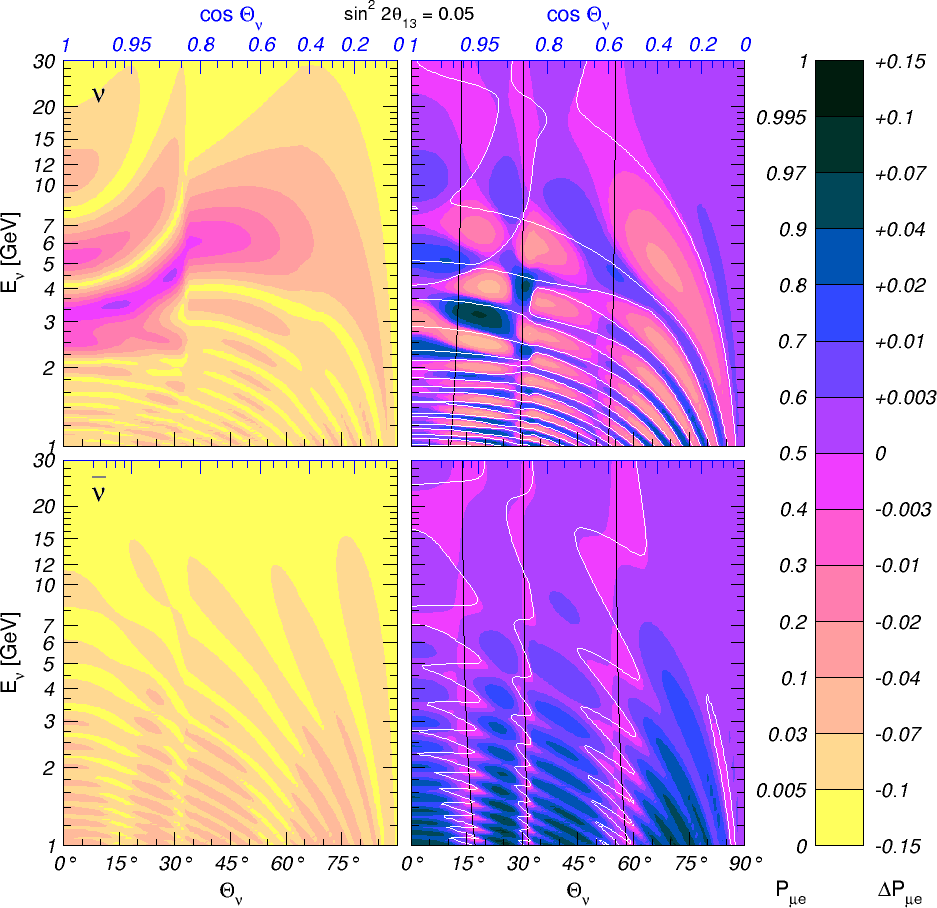}
  \caption{\label{fig:pme-N}%
    Oscillograms for the $\nu_\mu - \nu_e$ channel. Shown are the
    contours of constant probability $P_{\mu e}$ (left) as well as
    constant difference $\Delta P_{\mu e}$ of $3\nu$ and $2\nu$
    probabilities (right), for neutrinos (upper panels) and
    antineutrinos (lower panels). The oscillation parameters for
    $3\nu$ probabilities are $\sin^2 2\theta_{13} = 0.05$, $\Dmq_{21}
    = 8 \times 10^{-5}~\eVq$, $\tan^2\theta_{12} = 0.45$ and $\delta =
    0$. For the $2\nu$ probabilities we used $\Dmq = \Dmq_{31}$. In
    the right panels we also show the solar magic lines (black) and
    the lines which correspond to the condition
    \eqref{eq:doublecond}.}}

Fig.~\ref{fig:pme-N} illustrates the effect of inclusion of the 1-2
mixing and mass splitting on the $\nu_\mu - \nu_e$ and $\bar{\nu}_\mu
- \bar{\nu}_e$ oscillation patterns for $E_\nu \geq 1~\GeV$.  The left
panels correspond to the $3\nu$ case, whereas the right ones show the
oscillograms for the difference of probabilities
\begin{equation}
    \Delta P_{\mu e} \equiv P_{\mu e} - P_{\mu e}(\Dmq_{21} = 0)
\end{equation}
and, similarly, for $P_{\bar{\mu} \bar{e}}$ and $\Delta P_{\bar{\mu}
\bar{e}}$. As in the $\nu_e - \nu_e$ case, this difference has two
contributions:
\begin{equation}
    \label{eq:demu2}
    \Delta P_{\mu e}
    \approx \Delta P_{\mu e}^S + \Delta P_{\mu e}^A \,.
\end{equation}
The solar amplitude contribution $\Delta P_{\mu e}^S$ is now dominated
by the interference term and turns out to be much larger than in the
$\nu_e - \nu_e$ case. The atmospheric contribution is determined by
the same $\Delta P^A = P_A - \mathring{P}_A$ as before, 
Eqs.~\eqref{eq:dpa} and \eqref{eq:dpa1}, with the additional factor
$s_{23}^2 \approx 0.5$:
\begin{equation}
    \Delta P_{\mu e} \approx  s_{23}^2 \Delta P^A
    + \sin 2 \theta_{23} \cos \theta_{13}^m
    A_A^\cst A_S^\cst \cos(\phi_{31}^m + \delta) \,.
\end{equation}

Now in the 1-3 resonance region, the interference is $\leq 0.03-0.04$,
whereas the atmospheric one $\leq 0.015$. Therefore, in the first
approximation it is $\Delta P_{\mu e}^S$ that determines the structure
of oscillograms for $\Delta P_{\mu e}$, whereas $\Delta P_{\mu e}^A$
leads to some corrections to this structure.

Using the factorization approximation we can write the solar
contribution as
\begin{equation}
    \label{eq:dmue-fact}
    \Delta P_{\mu e}^S \approx \cos \theta_{23} |A_S|
    \LT(2 \sin \theta_{23} |A_A| \cos (\phi_{31}^m + \delta)
    + \cos \theta_{23} |A_S| \RT) \,.
\end{equation}
This formula corresponds to the one in Eq.~\eqref{eq:Pmue1} and is
valid for the energies below the energy of the 1-3 resonance. In the
right panel of Fig.~\ref{fig:pme-N} we show the lines of vanishing
solar correction: $\Delta P_{\mu e}^S = 0$. This equality is satisfied
where either $|A_S| = 0$ (the solar magic lines), or 
\begin{equation}
    \label{eq:doublecond}
    2 \sin \theta_{23} |A_A| \cos (\phi_{31}^m + \delta)
    + \cos\theta_{23} |A_S| = 0
\end{equation}
(white lines). If the second term is neglected the equality
\eqref{eq:doublecond} splits into two conditions: $A_A = 0$ (the
atmospheric magic line, see Sec.~\ref{sec:magic}) and
$\cos(\phi_{31}^m + \delta) = 0$ (the interference phase condition
line). These lines form a grid which we will discuss in detail in
Sec.~\ref{sec:cpviol}. This grid describes rather well the domain
structure of the oscillograms. Some deviations of the actual domain 
borders from the grid are related to the corrections from $\Delta
P^A$. Representing $\Delta P^A$ in \eqref{eq:dpa} as $\Delta P^A =
A_A^\cst K$, we can rewrite $\Delta P_{\mu e}$ in the constant density
approximation in the following form (up to corrections
$\mathcal{O}(P_S)$):
\begin{equation}
    \Delta P_{\mu e} \approx A_A^\cst \LT[ s_{23}^2 K
    + \sin 2 \theta_{23} \cos\theta_{13}^m A_S^\cst
    \cos(\phi_{31}^m + \delta) \RT] \,.
\end{equation}
Thus, with the atmospheric term taken into account, the lines
$A_A^\cst = 0$ still determine zeros of the difference $\Delta P_{\mu
e}$. However, the other lines of the condition $\Delta P_{\mu e} = 0$
are shifted by the term $s_{23}^2 K$ with respect to the solar magic
and the interference phase lines.

Notice that the corrections to the $2\nu$ oscillograms are enhanced in
the 1-3 resonance region, especially in the core domain at $3-4~\GeV$,
where $A_A$ is large due to the parametric resonance. Below $2~\GeV$
the corrections increase with decreasing energy because $\Delta P_{\mu
e}^S \propto P^S \propto 1 / E_\nu$ and $\Delta P_{\mu e}^A \propto 1
/ E_\nu$.

One qualitatively new feature of the oscillograms with $\Dmq_{21} \ne
0$ is that the absolute minima of $P_{\mu e}$ appear there as isolated
points (see Fig.~\ref{fig:pme-N}). In contrast, in the limit
$\Dmq_{21} = 0$ the absolute minima of $P_{\mu e}$ (and
$P_{\bar{\mu}\bar{e}}$) never appear as isolated points in the
oscillograms, but always form continuous curves (valleys of zero
probability).  This is unlike for the absolute maxima, such as the MSW
mantle peak or the parametric resonance peak in the core region, where
even in the limit of zero 1-2 splitting the value $P_A = 1$ (\ie,
$P_{\mu e} = s_{23}^2 P_A = s_{23}^2$) is reached only at a few
isolated points. This feature is a consequence of the symmetry of the
matter density profile of the Earth, and can be readily understood in
the following way.

The condition for the absolute minimum of the transition probability,
$P_{\mu e} = 0$, or $S_{e\mu} = 0$, where $S$ is the evolution matrix,
can be written as
\begin{equation}
    \label{eq:abmin}
    \Re(S_{e\mu}) = 0\,, \qquad \Im(S_{e\mu}) = 0 \,,
\end{equation}
and for a generic profile the absolute minima of $P_{\mu e}$ are found
as the points where the curves corresponding to the two conditions
in~\eqref{eq:abmin} intersect. However, due to the symmetry of the
Earth's matter density profile, in the 2-flavor approximation
($\Dmq_{21} = 0$) in the basis where the effective 2-flavor
Hamiltonian of the system is traceless the transition amplitude is
pure imaginary~\cite{Akhmedov:2001kd}. This means that the condition
$\Re[S_{e\mu}] = 0$ is satisfied automatically for all values of
$E_\nu$ and $\Theta_\nu$. Therefore, the zeros of $P_{\mu e}$ simply
coincide with the contour curves $\Im[S_{e\mu}] = 0$. For $\Dmq_{21}
\ne 0$ this is no longer the case, because the 2-flavor approximation
does not apply.

For antineutrinos (lower panels of Fig.~\ref{fig:pme-N}) the
corrections are again determined by the interference term with
somewhat smaller atmospheric contribution. Therefore, one can see a
domain structure with vertical lines. With the decrease of energy the
maxima of the corrections inside the domains monotonically increase,
since so do both the solar and atmospheric amplitudes. Notice also
that for $\delta = 0$ the positive corrections are larger than
negative; the situation in the neutrino channel is opposite.

As we pointed out in Sec.~\ref{sec:3fosc}, for the inverse channel one
has $P_{e\mu} = P_{\mu e} (\delta \to -\delta)$, where it has been
taken into account that the Earth density profile is symmetric.

According to Eqs.~\eqref{eq:Pmue} and \eqref{eq:Ptaue} the oscillation
probabilities $P_{\tau e}$ and $P_{e\tau}$ can be obtained from the
corresponding probabilities $P_{\mu e}$ and $P_{e\mu}$ through the
substitution $s_{23} \to c_{23}$, $c_{23} \to
-s_{23}$~\cite{Akhmedov:2004ny}. The interference term has the
opposite signs for channels including $\nu_\tau$ as compared with
those with $\nu_{\mu}$, which can be obtained from the unitarity
condition $P_{ee} + P_{\mu e} + P_{\tau e} = 1$ and the fact that
$P_{ee}$ does not depend on $\delta$.


\subsection{$\nu_\mu - \nu_\mu$ and $\nu_\tau - \nu_\tau$ channels}

The $\nu_\mu$ survival probability, $P_{\mu\mu}$, for symmetric
density profiles is given in Eq.~\eqref{eq:Pmumu}. It can be rewritten
as
\begin{multline}
    \label{eq:pmumutot}
    P_{\mu\mu} =
    |c_{23}^2 A_{\tilde{2}\tilde{2}} + s_{23}^2 A_{\tilde{3}\tilde{3}}|^2
    \\
    + 2\sin 2\theta_{23} \cos\delta\, \Re \LT[
    A_{\tilde{2}\tilde{3}}^* (c_{23}^2 A_{\tilde{2}\tilde{2}}
    + s_{23}^2 A_{\tilde{3}\tilde{3}}) \RT]
    + \sin^2 2\theta_{23} \cos^2\delta |A_{\tilde{2}\tilde{3}}|^2 \,.
\end{multline}
Note that $P_{\mu\mu}$ is an even function of $\delta$. Since
$A_{\tilde{2}\tilde{3}} = \mathcal{O}(r_\Delta s_{13})$ is a small
quantity, one can to a very good approximation neglect the term $\sim
\cos^2\delta$ in Eq.~\eqref{eq:Pmumu}, which is proportional to
$|A_{\tilde{2}\tilde{3}}|^2$. The term $\sim \cos^2\delta$ can only
become important when the main terms in $P_{\mu\mu}$ are small;
however, this happens only in very small regions of the experimental
parameter space.

In the limit $\Dmq_{21} \to 0$ we have $A_{\tilde{2}\tilde{2}} = 1$,
$A_{\tilde{2}\tilde{3}} = 0$. Then, parametrizing the 33-amplitude as
\begin{equation}
    \label{eq:parama33}
    A_{\tilde{3}\tilde{3}} \equiv
    |A_{\tilde{3}\tilde{3}}| e^{-i 2\phi_{\tilde{3}\tilde{3}}^m} =
    \sqrt{1 - P_A} e^{-i 2\phi_{\tilde{3}\tilde{3}}^m}
\end{equation}
we obtain from \eqref{eq:pmumutot}
\begin{equation}
    \label{eq:mm-lim}
    P_{\mu\mu} (\Dmq_{21} = 0)
    = 1 - \sin^2 2\theta_{23} \sin^2 \phi_{\tilde{3}\tilde{3}}^m
    - s_{23}^2 P_A - \sin^2 2\theta_{23}
    \cos 2\phi_{\tilde{3}\tilde{3}}^m (1 - \sqrt{1 - P_A}) \,.
\end{equation}
If in addition $\theta_{13} = 0$, then $P_A = 0$ and $P_{\mu\mu}
(\Dmq_{21} = 0)$ is reduced to the standard $2\nu$ vacuum oscillation
probability with $\phi_{\tilde{3}\tilde{3}}^m = \phi_A^0$.

It is easy to estimate the effect of the 1-2 mixing in the limit
$\theta_{13} = 0$. In this case the eigenstate $\tilde{\nu}_3$
decouples in the propagation basis, $A_{\tilde{2}\tilde{3}} = 0$, and
the probability takes a very simple form
\begin{equation}
    \label{eq:pmumu00}
    P_{\mu\mu} =
    |c_{23}^2 A_{\tilde{2}\tilde{2}} + s_{23}^2 A_{\tilde{3}\tilde{3}}|^2.
\end{equation}
Now
\begin{equation}
    A_{\tilde{3}\tilde{3}} = e^{-i 2 \phi_{\tilde{3}\tilde{3}}^0},
    \qquad \phi_{\tilde{3}\tilde{3}}^0 = \frac{\Dmq_{31} L}{4E_\nu} \,,
\end{equation}
and parametrizing $A_{\tilde{2}\tilde{2}}$ as
\begin{equation}
    A_{\tilde{2}\tilde{2}}
    = |A_{\tilde{2}\tilde{2}}| e^{-i 2\phi_{\tilde{2}\tilde{2}}^m} =
    \sqrt{1 - P_S} e^{-i 2\phi_{\tilde{2}\tilde{2}}^m} \,,
\end{equation}
we obtain
\begin{equation}
    \label{eq:mm-lim2}
    P_{\mu\mu} = 1
    - \sin^2 2\theta_{23} \sin^2 (\phi_{\tilde{3}\tilde{3}}^0
    - \phi_{\tilde{2}\tilde{2}}^m) - c_{23}^4 P_S
    - \frac{1}{2}\sin^2 2\theta_{23}
    \cos 2(\phi_{\tilde{3}\tilde{3}}^0 - \phi_{\tilde{2}\tilde{2}}^m)
    \LT(1 - \sqrt{1 - P_S}\RT).
\end{equation}
At high energies, $P_S \ll 1$, and consequently Eq.~\eqref{eq:mm-lim2}
becomes
\begin{equation}
    P_{\mu\mu} = P_{\mu\mu}^{2\nu} -
    P_S c_{23}^2 [c_{23}^2 + s_{23}^2 
    \cos 2(\phi_{\tilde{3}\tilde{3}}^0 - \phi_{\tilde{2}\tilde{2}}^m)] \,,
\end{equation}
where $P_{\mu\mu}^{2\nu} \equiv 1 - \sin^2 2\theta_{23} \sin^2
(\phi_{\tilde{3}\tilde{3}}^m - \phi_{\tilde{2}\tilde{2}}^m)$. In the
constant density approximation and above the 1-2 resonance we have
from \eqref{eq:ample22} $\phi_{\tilde{2}\tilde{2}}^m = \phi_{21}^m$,
and therefore $\phi_{\tilde{3}\tilde{3}}^m -
\phi_{\tilde{2}\tilde{2}}^m = \phi_{32}^m$. Thus, the effect of the
1-2 mixing is reduced to a shift of the oscillation phase and small
additive correction of order $P_S$. Larger corrections are expected
for non-zero 1-3 mixing due to the interference of the 1-2 and 1-3
modes.

\FIGURE[!t]{
  \includegraphics[width=146mm]{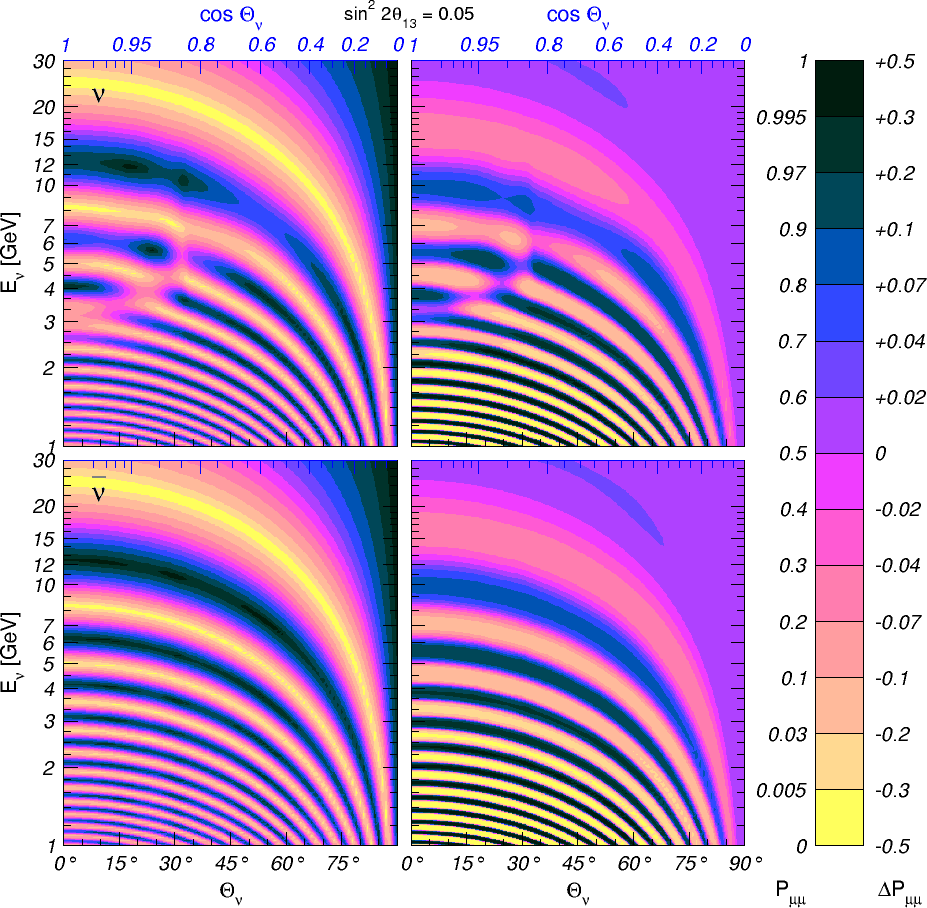}
  \caption{\label{fig:pmm-N}%
    Oscillograms for the $\nu_\mu - \nu_\mu$ channel. Shown are the
    contours of constant probability $P_{\mu\mu}$ (left) as well as
    constant difference $\Delta P_{\mu\mu}$ of $3\nu$ and $2\nu$
    probabilities (right), for neutrinos (upper panels) and
    antineutrinos (lower panels). The oscillation parameters for
    $3\nu$ probabilities are $\sin^2 2\theta_{13} = 0.05$, $\Dmq_{21}
    = 8 \times 10^{-5}~\eVq$, $\tan^2\theta_{12} = 0.45$ and $\delta =
    0$. For the $2\nu$ probabilities we used $\Dmq = \Dmq_{31}$.}}

In Fig.~\ref{fig:pmm-N} we show the oscillograms for $P_{\mu\mu}$ and
$P_{\bar{\mu}\bar{\mu}}$ (left upper and lower panels) and the
differences $\Delta P_{\mu\mu} \equiv P_{\mu \mu} - P_{\mu\mu}
(\Dmq_{21} = 0)$ and $\Delta P_{\bar{\mu}\bar{\mu}} \equiv
P_{\bar{\mu}\bar{\mu}} - P_{\bar{\mu}\bar{\mu}} (\Dmq_{21} = 0)$
(right panels).  One can see in this figure the regular oscillatory
pattern dominated by the vacuum $\nu_\mu\to \nu_\tau$ oscillations
with certain distortion in the region of 1-3 the resonances: the MSW
resonance in the mantle and core and the parametric ridges in the core
domain. The dominant effect of the 1-2 mixing is related to the phase
shift of this main mode, which increases with decreasing energy: at
low energies the corrections become of order 1. $\Delta P_{\mu\mu}$
follows to a large extend the structure of $P_{\mu\mu}$. No domain
structure appears here. All these features can be seen from the
formulas \eqref{eq:pmumutot}, \eqref{eq:mm-lim}. In particular,
according to \eqref{eq:mm-lim2}, for zero 1-3 mixing we obtain
\begin{equation}
    \label{eq:lim13zero}
    \Delta P_{\mu\mu} = \sin^2 2\theta_{23} \LT[
    \sin^2(\phi_{\tilde{3}\tilde{3}}^0 - \phi_{\tilde{2}\tilde{2}}^m)
    - \sin^2 \phi_{\tilde{3}\tilde{3}}^0 \RT]
    + \mathcal{O}(P_S) \,.
\end{equation}

An additional insight can be gained using the constant density
approximation. Let us consider the case of maximal 2-3 mixing for
which the figures have been plotted. For maximal mixing and zero phase
$\delta$ the interference term in the $\nu_\mu - \nu_\mu$ channel
coincides, up to the sign, with the one in the $\nu_\mu - \nu_e$
channel (see Sec.~\ref{sec:cpviol}), which has been estimated in the
previous subsection.  Therefore, let us now consider the first term in
\eqref{eq:pmumutot}. In the limit $\Dmq_{21} \to 0$ we obtain for this
term a very simple expression:
\begin{equation}
    \label{eq:mm-lim3}
    P_{\mu\mu} (\Dmq_{21} = 0)
    = 1 - (1 - \sin^4 \theta_{13}^m) \sin^2 \mathring{\phi}_{32}^m \,.
\end{equation}
Here the phase $\mathring{\phi}_{32}^m$ should be calculated in the
$2\nu$ context with $H_2^m = 0$.

The survival probabilities $P_{\tau\tau}$ and
$P_{\bar{\tau}\bar{\tau}}$ can be obtained from the corresponding
probabilities $P_{\mu\mu}$ and $P_{\bar{\mu}\bar{\mu}}$ through the
substitution $s_{23} \to c_{23}$, $c_{23} \to
-s_{23}$~\cite{Akhmedov:2004ny}.


\subsection{$\nu_\mu - \nu_\tau$ channel}

The probability of $\nu_\mu \to \nu_\tau$ oscillations for symmetric
matter density profiles is given in \eqref{eq:Pmutau}. It can be
rewritten as
\begin{multline}
    \label{eq:mmtt}
    P_{\mu\tau} = \frac{1}{4}\sin^2 2\theta_{23} |A_{\tilde{2}\tilde{2}}
    - A_{\tilde{3}\tilde{3}}|^2
    + \sin 2\theta_{23} \cos 2\theta_{23} 
    \cos\delta\, \Re\LT[ (A_{\tilde{3}\tilde{3}}^* - A_{\tilde{2}\tilde{2}}^*)
    A_{\tilde{2}\tilde{3}} \RT]
    \\
    - \sin 2\theta_{23} \sin\delta \,
    \Im\LT[ A_{e\tilde{2}}^* A_{e\tilde{3}} \RT]
    + (1 - \sin^2 2\theta_{23} \cos^2 \delta) |A_{\tilde{2}\tilde{3}}|^2 \,.
\end{multline}
The oscillations are mainly driven by $\Dmq_{31}$ and the large mixing
angle $\theta_{23}$. The amplitude depends on $\delta$ through the
terms proportional to $\cos\delta$ and $\sin\delta$, and therefore
$P_{\mu\tau}$ contains both CP- and T-even and odd terms. Due to
unitarity, all the information on $\nu_\mu\to\nu_\tau$ oscillations is
contained in the already discussed probabilities $P_{\mu\mu}$ and
$P_{\mu e}$: $P_{\mu\tau} = 1 - P_{\mu\mu} - P_{\mu e}$. Furthermore,
one can show that the $\delta$-dependent interference terms
proportional to $\sin\delta$ and $\cos\delta$ satisfy the following
relation
\begin{equation}
    P_{\mu\tau}^{\delta} = - P_{\mu e}^{\delta} - P_{\mu\mu}^{\delta}
\end{equation}
(see the next section for details).

Notice that for the maximal 2-3 mixing and $\delta = 0$, the
probability takes a very simple form
\begin{equation}
    P_{\mu\tau} = \frac{1}{4}
    |A_{\tilde{2}\tilde{2}} - A_{\tilde{3}\tilde{3}}|^2 \,.
\end{equation}
In the limit $\Dmq_{21} \to 0$ we obtain
\begin{equation}
    P_{\mu\tau}(\Dmq_{21} = 0)  =
    \frac{1}{4} |1 - A_{\tilde{3}\tilde{3}}|^2  =
    \frac{1}{4} \LT( 2 - P_A - 2 \sqrt{1- P_A}
    \cos 2\phi_{\tilde{3}\tilde{3}}^m \RT),
\end{equation}
where $\phi_{\tilde{3}\tilde{3}}^m$ was defined in
\eqref{eq:parama33}. For small $P_A$ the probability becomes
\begin{equation}
    P_{\mu\tau} (\Dmq_{21} = 0) =
    \LT( 1 - \frac{1}{2} P_A \RT) \sin^2 \phi_{32}^m \,,
\end{equation}
which in turn reduces to the standard $2\nu$ probability for
$\theta_{13} = 0$.

The oscillograms for the $\nu_\mu - \nu_\tau$ channel are very similar
to those for the $\nu_\mu - \nu_\mu$ channel plotted in
Fig.~\ref{fig:pmm-N}. To a large extent they are complementary in the
sense that the corresponding minima and maxima are interchanged. The
$P_{\mu\tau}$ oscillograms exhibit the vacuum oscillations pattern
everywhere apart from the region $E_\nu \simeq 3 - 12~\GeV$. In this
region the pattern is distorted by the 1-3 level crossing, as well as
by the parametric enhancement of the oscillations in the 1-3 mode.
This distortion is absent in the antineutrino channel. As in the
$\nu_\mu - \nu_\mu$ case, the difference of the probabilities, $\Delta
P_{\mu\tau}$, is dominated by the phase shift, and the corrections
have the opposite sign compared to $\Delta P_{\mu\mu}$.

Let us now present some results for constant density matter which will
allow us to quantify the features described above.  Using the
expressions for the amplitudes in Eqs.~\eqref{eq:ample22} and
\eqref{eq:ample33}, we find from Eq.~\eqref{eq:mmtt} for $E_\nu \gg
E_{12}^R$ (the maximal 2-3 mixing and $\delta = 0$) 
\begin{equation}\begin{split}
    \label{eq:pmutauc}
    P_{\mu\tau}^\cst &= \LT|
    e^{-i\phi_{31}^m} \cos^2 \theta_{13}^m \sin \phi_{32}^m +
    \LT[ 1 - \cos^2 \theta_{12}^m (1 + \sin^2 \theta_{13}^m) \RT]
    \sin\phi_{21}^m \RT|^2
    \\
    & \approx \LT|e^{-i\phi_{31}^m}  \cos^2 \theta_{13}^m
    \sin \phi_{32}^m + \sin \phi_{21}^m \RT|^2 \,.
\end{split}\end{equation}
For energies above the 1-3 resonance one, $\cos^2 \theta_{13}^m \to 0$
and Eq.~\eqref{eq:pmutauc} gives $P_{\mu\tau} \approx \sin^2
\phi_{21}^m$ with $\phi_{21}^m \approx \Dmq_{31} L/2E_\nu$. Therefore
the corrections to $P_{\mu\tau}$ due to the vacuum 1-2 mixing and
splitting are strongly suppressed. For energies between the two
resonances, using the phase exchange relation \eqref{eq:phaseexc}, we
obtain
\begin{equation}
    P_{\mu\tau}^\cst \approx
    \cos^4 \theta_{13}^m \sin^2 \phi_{31}^m +
    \frac{1}{2} \sin^2 2\theta_{13}^m \sin\phi_{21}^m
    \sin\phi_{31}^m \cos\phi_{32}^m
    + \sin^4 \theta_{13}^m \sin^2 \phi_{21}^m \,.
\end{equation}
For $\Dmq_{21} = 0$ we have
\begin{multline}
    P_{\mu\tau}^\cst (\Dmq_{21} = 0) \approx
    \cos^4 \theta_{13}^m \sin^2 \phi_{32}^m
    \\
    - \frac{1}{2} \sin^2 2\theta_{13}^m \sin\phi_{21}^m
    \sin\phi_{31}^m \cos\phi_{32}^m
    + \sin^4 \theta_{13}^m \sin^2 \phi_{21}^m \,,
\end{multline}
where the mixing angle $\theta_{13}^m$ and the phases should be
calculated in the $2\nu$ context. From the level crossing scheme for
the normal mass hierarchy we obtain for energies below the 1-3
resonance one that $\mathring{\phi}_{21}^m \approx -\phi_{21}^m$ and
$\mathring{\phi}_{32}^m \approx \phi_{31}^m$, and the difference between
the $2\nu$ and $3\nu$ phases is proportional to $\Dmq_{21} L/2E_\nu$.
This difference increases with decreasing energy and at $E_\nu \sim
1~\GeV$ it can be of order $\pi/2$.


\subsection{Inverted mass hierarchy}
\label{sec:inverted}

Let us briefly comment on the features of the oscillograms for the
inverted mass hierarchy, \ie, for $\Dmq_{31} < 0$. The main change as
compared to the normal hierarchy is due to the 1-3 resonance structure
which appears in the antineutrino channel now. Pulling out of the
brackets in the Hamiltonian \eqref{eq:matr1} the positive factor
$|\Dmq_{31}|$, we have to change the sign in front of all the terms
without $r_\Delta$ and $V$ in the matrix \eqref{eq:matr1}. The general
oscillation formulas we presented before are valid in this case,
however the eigenvalues of the Hamiltonian and mixing angles in matter
should be changed. The level crossing scheme is also modified. In the
neutrino channel (where there is the 1-2 resonance only), in the limit
of large energies we have
\begin{equation}
    \label{eq:invhi}
    H_{1}^m \approx \frac{\Dmq_{21} c_{12}^2}{2E_\nu} \,, \qquad
    H_{2}^m \approx V \,, \qquad
    H_{3}^m \approx \frac{\Dmq_{31}}{2E_\nu} \,.
\end{equation}
In the antineutrino channel, below the 1-3 resonance the eigenvalues
are
\begin{equation}
    \label{eq:belowinv}
    H_{1}^m \approx V \,, \qquad
    H_{2}^m \approx \frac{\Dmq_{21} c_{12}^2}{2E_\nu} \,, \qquad
    H_{3}^m \approx \frac{\Dmq_{31} c_{13}^2}{2E_\nu} \,,
\end{equation}
while above the 1-3 resonance,
\begin{equation}
    \label{eq:belowab}
    H_{1}^m \approx \frac{\Dmq_{31} c_{13}^2}{2E_\nu} \,, \qquad
    H_{2}^m \approx \frac{\Dmq_{21} c_{12}^2}{2E_\nu} \,, \qquad
    H_{3}^m \approx V \,.
\end{equation}

\FIGURE[!t]{
  \includegraphics[width=147mm]{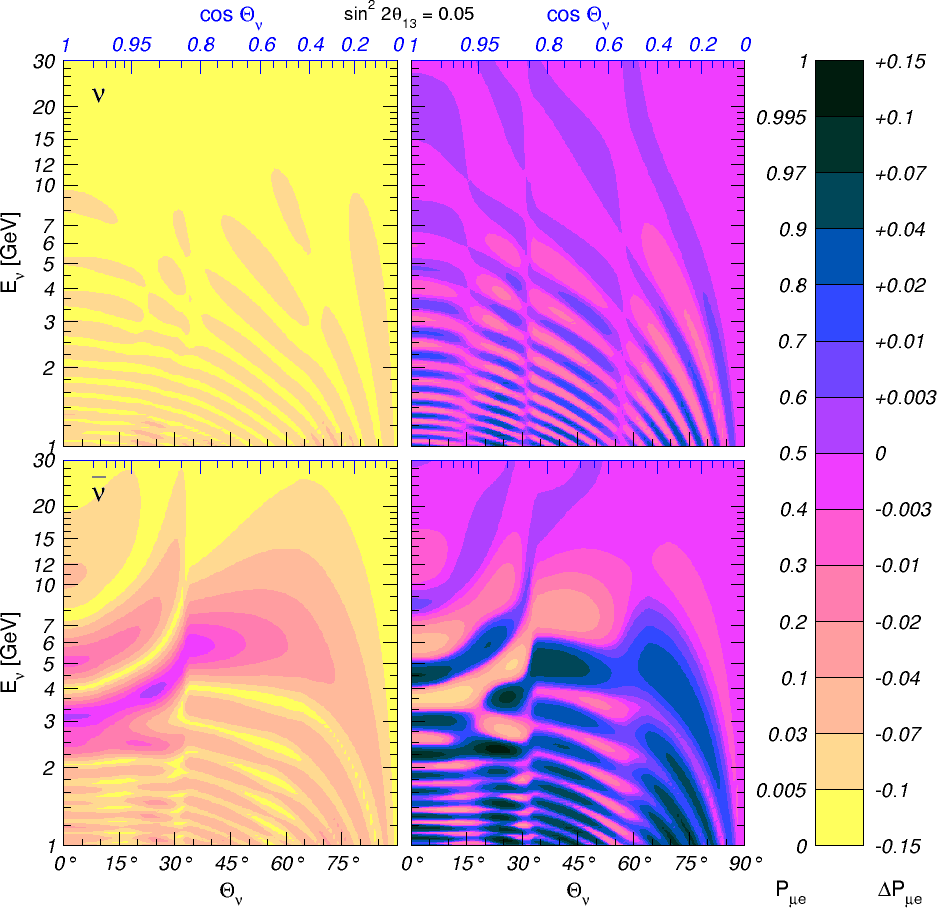}
  \caption{\label{fig:pme-I}%
    Oscillograms for the $\nu_\mu - \nu_e$ channel in the case of the 
    inverted mass hierarchy. Shown are the contours of constant
    probability $P_{\mu e}$ (left) as well as constant difference
    $\Delta P_{\mu e}$ of $3\nu$ and $2\nu$ probabilities (right), for
    neutrinos (upper panels) and antineutrinos (lower panels). The
    oscillation parameters for $3\nu$ probabilities are $\sin^2
    2\theta_{13} = 0.05$, $\Dmq_{21} = 8 \times 10^{-5}~\eVq$,
    $\tan^2\theta_{12} = 0.45$ and $\delta = 0$. For the $2\nu$
    probabilities we used $\Dmq = \Dmq_{31}$.}}

Here for illustration we will consider the oscillograms for the
$\nu_\mu - \nu_e$ channel only. The other channels can be analyzed in
a similar way. In Fig.~\ref{fig:pme-I} we show the oscillograms for
$P_{\mu e}$ (left panels), and for the difference of $3\nu$ and $2\nu$
probabilities, $\Delta P_{\mu e}$ (right panels).

In the approximation of $\Dmq_{21} = 0$ the neutrino oscillograms for
the inverted hierarchy coincide with the antineutrino oscillograms for
the normal hierarchy, and vice-versa, provided that $\Dmq_{31}$ are
taken to be the same in both cases. The inclusion of the 1-2 mixing
and mass splitting breaks this symmetry. However, at high energies
where the corrections are small, the correspondence ``$\nu$ inverted
$\leftrightarrow$ $\bar\nu$ normal'', ``$\bar\nu$ inverted
$\leftrightarrow$ $\nu$ normal'' approximately holds. In our
computations in the limit $\Dmq_{21} \to 0$ we have taken the
remaining mass squared difference to be equal to the largest splitting
in the $3\nu$ context, $\Dmq = \Dmq_{32}$.

In the neutrino channel now there is only the 1-2 resonance and the
height of the oscillation peaks of the probability increases with
decreasing energy.  The difference of $3\nu$ and $2\nu$ probabilities
is mainly due to the interference terms, as in the normal mass
hierarchy case. This explains why one can clearly see a domain
structure with vertical lines. The solar magic lines are shifted due
to the contribution from $\Delta P^A$. At high energies negative
corrections are larger in the absolute value than the positive ones
(for $\delta = 0$). Notice also that the shape of domains for
neutrinos in the case of inverted hierarchy is similar to that for
antineutrinos in the normal mass hierarchy case.

In the antineutrino channel the oscillation pattern is dominated by
the MSW and parametric resonances. The difference of probabilities in
the $3\nu$ and $2\nu$ contexts is again due to the interference term
$\propto A_S A_A$, with some corrections from $\Delta P_A$. According
to the figure, the 1-2 mixing and splitting effect is enhanced in the
regions of the MSW resonance peaks and along the parametric ridges,
since $A_A$ is enhanced there. The oscillogram $\Delta P_{\mu e}$ has
a domain structure. The heights of the peaks are maximal in the
resonance regions. In the mantle domain the heights reach minimum at
$E_\nu \sim 3~\GeV$ and in the core domain at $E_\nu \sim 2~\GeV$,
similarly to what we had for the CP-sensitivity peaks
(Sec.~\ref{sec:cpviol}). Due to the interplay of the interference
terms and $\Delta P^A$, for $\delta = 0$ the corrections in the case
of the inverted mass hierarchy are larger than those in the normal
mass hierarchy case: \eg, in the region of the 1-3 resonance we have
$\Delta P_{\mu e} \sim \pm (0.07 - 0.10)$. Below $3 - 4~\GeV$ the
positive corrections dominate.

One can use the analytic formulas of Sec.~\ref{sec:12split} with
appropriately changed phases and mixing angles to describe these
results quantitatively.


\section{Effects of CP-violating phase $\delta$}
\label{sec:cpviol}

In this section we consider in detail the properties of the
CP-interference terms and in particular their dependence on the phase
$\delta$ in different channels.
As can be seen from the expressions given in Sec.~\ref{sec:3fosc}, the
survival probability $P_{ee}$ does not depend on the CP-violating
phase $\delta$, both for oscillations in vacuum and in
matter~\cite{Kuo:1987km, Minakata:1999ze}. This is a consequence of
the facts that (i) $\delta$ is rotated away by transforming to the
propagation basis, and (ii) the probability $P_{ee}$ is not affected
by this transformation. Note that for oscillations in vacuum or in
matter with symmetric density profiles, the other two survival
probabilities, $P_{\mu\mu}$ and $P_{\tau\tau}$, depend on $\delta$
only through the terms proportional to $\cos\delta$ and $\cos
2\delta$~\cite{Yokomakura:2002av} since they are T-even quantities. In
contrast to this, for oscillations in a matter whose density profile
is not symmetric with respect to the midpoint of the neutrino
trajectory, these probabilities acquire also terms proportional to
$\sin\delta$ and $\sin 2\delta$.


\subsection{Interference and CP-violation}
\label{sec:interf}

The unitarity of the evolution matrix in the propagation basis
\eqref{eq:matr2} for symmetric density profiles gives
\begin{equation}
    \label{eq:unit}
    A_{\tilde{2}\tilde{3}} A_{\tilde{2}\tilde{2}}^* + A_{\tilde{2}\tilde{3}}^*
    A_{\tilde{3}\tilde{3}} = -A_{e\tilde{2}}^* A_{e\tilde{3}} \,.
\end{equation}
This relation allows one to write explicitly the $\delta$-dependent
terms for different oscillation channels as
\begin{align}
    \label{eq:delta-me}
    P_{\mu e}^\delta &= \sin 2\theta_{23}
    \LT\lbrace
    \cos\delta \Re[ A_{e\tilde{2}}^* A_{e\tilde{3}} ] +
    \sin\delta \Im[A_{e\tilde{2}}^* A_{e\tilde{3}} ]
    \RT\rbrace,
    \\
    \label{eq:delta-mm}
    P_{\mu\mu}^\delta &= \sin 2\theta_{23} \cos\delta
    \LT\lbrace
    - \Re[ A_{e\tilde{2}}^* A_{e\tilde{3}} ] -
    \cos 2\theta_{23}\Re[ A_{\tilde{2}\tilde{3}}^*(A_{\tilde{3}\tilde{3}}
    - A_{\tilde{2}\tilde{2}}) ] \RT\rbrace,
    \\
    \label{eq:delta-mt}
    P_{\mu\tau}^\delta &= \sin 2\theta_{23}
    \LT\lbrace
    - \sin\delta \Im[A_{e\tilde{2}}^* A_{e\tilde{3}}]
    + \cos\delta \cos 2\theta_{23} \Re[ A_{\tilde{2}\tilde{3}}^*
    (A_{\tilde{3}\tilde{3}} - A_{\tilde{2}\tilde{2}}) ]
    \RT\rbrace.
\end{align}
Here in $P_{\mu\mu}^\delta$ and $P_{\mu\tau}^\delta$ we have omitted
small terms proportional to $|A_{\tilde{2}\tilde{3}}|^2$. The sum of
these interference terms is zero. For maximal 2-3 mixing
Eqs.~\eqref{eq:delta-me}, \eqref{eq:delta-mm} and \eqref{eq:delta-mt}
reduce to
\begin{align}
    P_{\mu e}^\delta
    &= \cos\delta \Re[ A_{e\tilde{2}}^* A_{e\tilde{3}} ]
    + \sin\delta \Im[ A_{e\tilde{2}}^* A_{e\tilde{3}} ] \,,
    \\
    P_{\mu\mu}^\delta
    &= -\cos\delta \Re[ A_{e\tilde{2}}^* A_{e\tilde{3}} ] \,,
    \\
    P_{\mu\tau}^\delta
    &= -\sin\delta \Im[ A_{e\tilde{2}}^* A_{e\tilde{3}} ] \,.
\end{align}
The following consequences of these expressions can be useful for
measurements of $\delta$: if $\delta = 0$, the probability
$P_{\mu\tau}$ does not contain the interference term and
$P_{\mu\mu}^\delta = -P_{\mu e}^\delta$; if $\delta = \pi/2$,
$P_{\mu\mu}$ has no interference term and $P_{\mu\tau}^\delta =
-P_{\mu e}^\delta$.

Using the phase $\phi \equiv \arg(A_{e\tilde{2}}^* A_{e\tilde{3}})$
defined in Eq.~\eqref{eq:arg} we obtain in the general case
\begin{align}
    \label{eq:Pme-2}
    P_{\mu e}^\delta &= \hphantom{-} \sin 2\theta_{23}
    \cos(\phi - \delta) \, |A_{e\tilde{2}} A_{e\tilde{3}}| \,,
    \\
    P_{\mu\mu}^\delta &= -\sin 2\theta_{23} \cos\delta \cos\phi \,
    |A_{e\tilde{2}} A_{e\tilde{3}}| - D_{23} \,,
    \\
    P_{\mu\tau}^\delta & = -\sin 2\theta_{23} \sin\delta \sin\phi \,
    |A_{e\tilde{2}} A_{e\tilde{3}}| + D_{23} \,,
\end{align}
where
\begin{equation}
    D_{23} \equiv \frac{1}{2} \sin 4\theta_{23} \cos\delta~
    \Re\LT[ A_{\tilde{2}\tilde{3}}^* (A_{\tilde{3}\tilde{3}}
    - A_{\tilde{2}\tilde{2}}) \RT]
\end{equation}
is proportional to the deviation of the 2-3 mixing from the maximal
one. Notice that $D_{23}$ enters into $P_{\mu\mu}^\delta$ and
$P_{\mu\tau}^\delta$ with opposite signs while $P_{\mu e}^\delta$ does
not depend on $D_{23}$ at all. $D_{23}$ is CP-even. It can be
estimated using the constant density approximation as
\begin{equation}
    D_{23}^\cst \approx -\frac{1}{2} \sin 4\theta_{23} \cos\delta
    \LT[ \cos\phi_{31}^m c_{13}^m A_A^\cst A_S^\cst - 2 \sin\theta_{13}^m
    \sin\phi_{21}^m A_S^\cst \RT] \,.
\end{equation}
This expression shows that corrections to the term proportional to
$A_A^\cst A_S^\cst$ in $D_{23}$ are in general not small.

For maximal 2-3 mixing one has
\begin{align}
    \label{eq:muemax}
    P_{\mu e}^\delta &= \hphantom{-}
    \cos(\phi - \delta) \, |A_{e\tilde{2}} A_{e\tilde{3}}| \,,
    \\
    \label{eq:mumumax}
    P_{\mu\mu}^\delta &= -\cos\delta \cos\phi \,
    |A_{e\tilde{2}} A_{e\tilde{3}}| \,,
    \\
    \label{eq:mutaumax}
    P_{\mu\tau}^\delta &= -\sin\delta \sin\phi \,
    |A_{e\tilde{2}} A_{e\tilde{3}}| \,.
\end{align}

The other $\delta$-dependent terms in the probabilities, which are
proportional to the square of the small quantity
$|A_{\tilde{2}\tilde{3}}|^2$, are
\begin{equation}
    P_{\mu\mu}^{\delta\delta} \equiv \cos^2\delta
    \sin^2 2\theta_{23} |A_{\tilde{2}\tilde{3}}|^2 \,,
    \qquad
    P_{\mu\tau}^{\delta\delta} \equiv (1 - \cos^2\delta
    \sin^2 2\theta_{23}) \, |A_{\tilde{2}\tilde{3}}|^2 \,.
\end{equation}
Notice that the sum of these terms does not depend on $\delta$ and
equals $|A_{\tilde{2}\tilde{3}}|^2$.

Let us present also the $\delta-$ dependent terms of the probabilities
for other channels. As we have already mentioned, $P_{\tau e}^\delta =
- P_{\mu e}^\delta$, and for the reverse channels, according to
\eqref{eq:Trev} we obtain
\begin{align}
    P_{e\mu}^\delta &= \sin 2\theta_{23} \cos(\phi + \delta) \,
    |A_{e\tilde{2}} A_{e\tilde{3}}| \,,
    \\
    P_{\tau\mu}^\delta &= \sin 2\theta_{23} \sin\delta \sin\phi \,
    |A_{e\tilde{2}} A_{e\tilde{3}}| + D_{23} \,.
\end{align}
As was pointed out above, $P_{\tau\tau}$ can be obtained from
$P_{\mu\mu}$ through the substitution $s_{23}\to c_{23}$, $c_{23}\to
-s_{23}$:
\begin{equation}
    P_{\tau\tau}^\delta = \sin 2\theta_{23} \cos\delta \cos\phi \,
    |A_{e\tilde{2}} A_{e\tilde{3}}| - D_{23} \,.
\end{equation}
For antineutrinos, according to \eqref{eq:pranti}, the probabilities
have the same form as the corresponding probabilities derived above
with the changed sign of $\delta$ and the amplitudes computed for the
opposite sign of the potential.

Thus, the $\delta$ dependent terms in all channels are expressed in
terms of two combinations of the propagation basis amplitudes,
$|A_{e\tilde{2}} A_{e\tilde{3}}|$ and $D_{23}$. In the case of the
maximal 2-3 mixing ($D_{23} = 0$), only the first combination enters
the interference terms. Furthermore, for the channels involving
electron neutrinos, $\nu_\mu - \nu_e$ and $\nu_\tau - \nu_e$, only the
first combination is relevant, even for the non-maximal 2-3 mixing.
For generic values of $\delta$ the CP-dependent terms in all the
channels but $\nu_e - \nu_e$ are of the same order.

To assess the $\delta$-dependent interference terms, one can consider
the difference of the oscillation probabilities for two different
values of the CP-phase:
\begin{equation}
    \Delta P_{\alpha\beta}^\text{CP}(\delta)
    \equiv P_{\alpha\beta}(\delta) - P_{\alpha\beta}(\delta_0) \,.
\end{equation}
In practice this would correspond to fit of the probability with the
true value of the phase $\delta$ by the probability with some assumed
value of the phase $\delta_0$.  In Figs.~\ref{fig:dcp-me},
\ref{fig:dcp-mm} and~\ref{fig:dcp-nomax} we show some examples of the
oscillograms for $\Delta P_{\mu e}^\text{CP}$ and $\Delta
P_{\mu\mu}^\text{CP}$ for different values of $\delta$ and $\delta_0 =
0^\circ$.  Although the CP-oscillograms appear to have a complex
structure, this structure can be readily understood in terms of the
three grids of curves, which we consider next.


\subsection{``Magic'' lines and interference phase lines}
\label{sec:magic}

Let us first consider the $\nu_\mu \to \nu_e$ oscillation probability,
for which the equality
\begin{equation}
    \label{eq:diffpro}
    \Delta P_{\mu e}^\text{CP}(\delta)
    \equiv P_{\mu e}(\delta) - P_{\mu e}(\delta_0)
    = P_{\mu e}^\delta(\delta) - P_{\mu e}^\delta(\delta_0)
\end{equation}
is exact. The condition $\Delta P_{\mu e}^\text{CP} = 0$ is equivalent
to
\begin{equation}
    \label{eq:fact}
    |A_{e\tilde{2}} A_{e\tilde{3}}| \cos(\phi - \delta)
    = |A_{e\tilde{2}} A_{e\tilde{3}}| \cos(\phi - \delta_0) \,.
\end{equation}
The same condition holds for the $\nu_e - \nu_\tau$ channel.
This equality is satisfied if at least one of the following three
conditions is fulfilled
\begin{equation}\begin{aligned}
    \label{eq:abc}
    \text{(A)} & \quad &
    A_{e\tilde{2}}(E_\nu, \Theta_\nu) &= 0 \,,
    \\
    \text{(B)} & \quad &
    A_{e\tilde{3}}(E_\nu, \Theta_\nu) &= 0 \,,
    \\
    \text{(C)} & \quad &
    \phi(E_\nu, \Theta_\nu) - \delta_0 &=
    - \LT[ \phi(E_\nu, \Theta_\nu) - \delta \RT] + 2\pi l \,.
\end{aligned}\end{equation}
The last condition implies
\begin{equation}
    \label{eq:C}
    \phi(E_\nu, \Theta_\nu) = (\delta + \delta_0) / 2 + \pi l \,.
\end{equation}
Under conditions (A) and (B) the equality \eqref{eq:fact} is satisfied
identically for all values of $\delta$. In these cases the transition
probability does not depend on CP-phase. It is also satisfied
trivially if $\delta = \delta_0 + 2\pi n$, since the true and assumed
values of the phase coincide.\footnote{In Figs.~\ref{fig:dcp-me},
\ref{fig:dcp-mm} and~\ref{fig:dcp-nomax} we show the lines of the
condition (C) which correspond to certain values of phases, which we
denote by $(\delta + \delta_0)_\text{fig}$.  According to
\eqref{eq:muemax} the interference term vanishes if $\phi = \delta +
\pi/2 + \pi n$.  Comparing this last equality with~\eqref{eq:C}, we
find that along the lines of the condition (C) the interference term
vanishes for value of phase $\delta = - (\delta +
\delta_0)_\text{fig}/2 + \pi/2 + \pi n$.}

In general, the conditions (A) and (B) can be satisfied at isolated
points in the ($\Theta_\nu, E_\nu$) plane only. Indeed, in order for
$A_{e\tilde{3}}$ to vanish, both its real and imaginary parts must be
zero. The conditions $\Re A_{e\tilde{3}} = 0$ and $\Im A_{e\tilde{3}}
= 0$ each define a set of curves in the ($\Theta_\nu, E_\nu$) plane,
and the curves from one set can only intersect with those from the
other at isolated points. The same applies to the condition
$A_{e\tilde{2}} = 0$. In contrast to this, as we will discussed below,
in the factorization approximation both the conditions $A_{e\tilde{2}}
= A_S = 0$ and $A_{e\tilde{3}} = A_A = 0$ are fulfilled along certain
curves in the oscillograms. This happens because the amplitudes $A_S$
and $A_{A}$ take a 2-flavor form. In the bases where the corresponding
$2 \times 2$ Hamiltonians are traceless, both $A_A$ and $A_S$ are pure
imaginary because of the symmetry of the Earth's density
profile~\cite{Akhmedov:2001kd}. Therefore for $A_{S}$ and $A_{A}$ to
be zero, it is enough to require that their imaginary parts vanish.
So, instead of conditions (A) and (B) in \eqref{eq:abc} we will
consider equalities $A_{S} = 0$ and $A_{A} = 0$.

\PAGEFIGURE{
  \includegraphics[width=143mm]{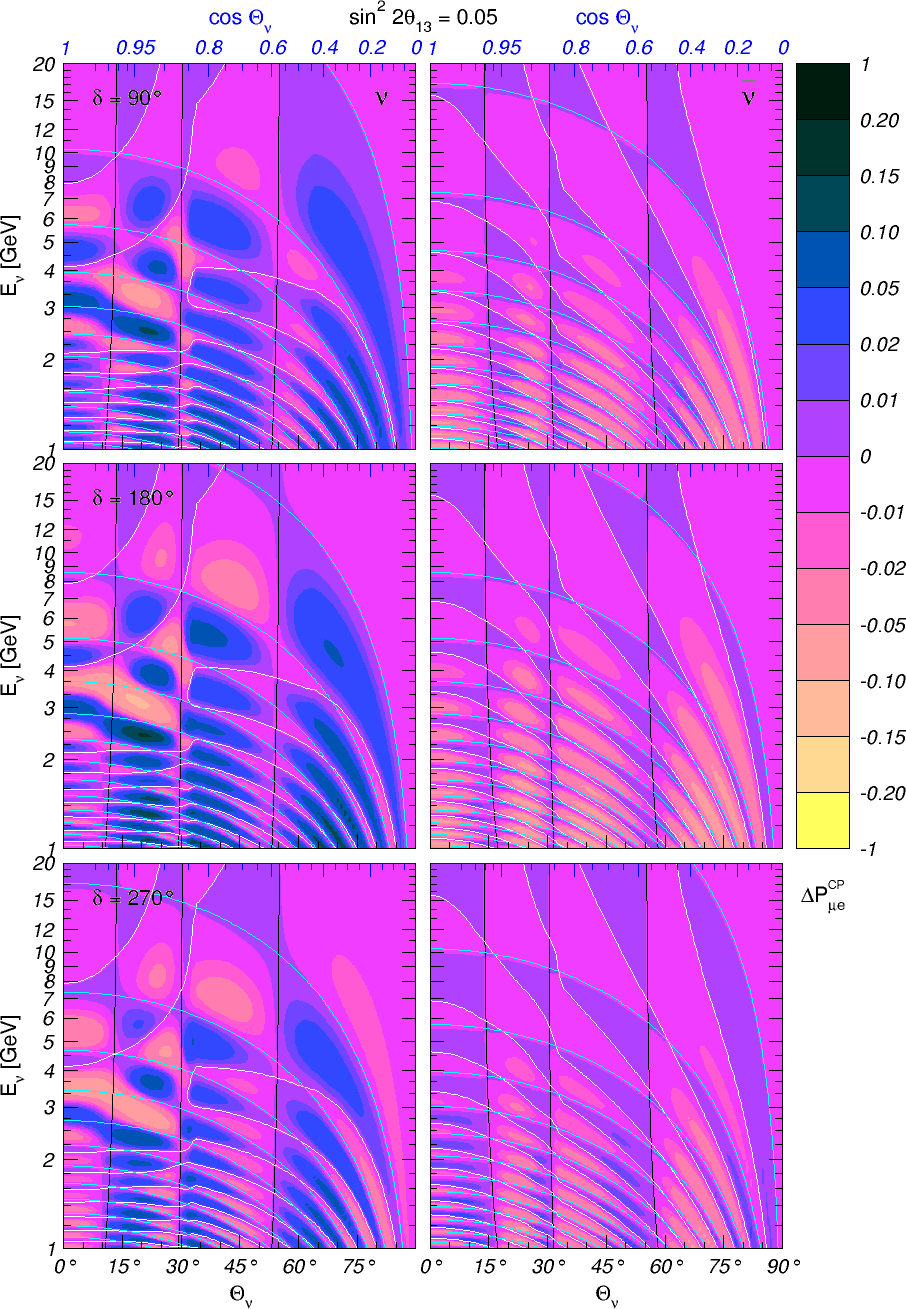}
  \caption{\label{fig:dcp-me}%
    Oscillograms for the difference of probabilities $\Delta P_{\mu
    e}^\text{CP}(\delta) = P_{\mu e}(\delta) - P_{\mu e}(\delta_0)$
    with $\delta_0 = 0^\circ$. Shown are the solar (black),
    atmospheric (white) and interference phase condition (cyan)
    curves.}}

In the factorization approximation the conditions in (A), (B) and (C)
define three sets of curves in the oscillograms (see
Fig.~\ref{fig:dcp-me}), which play crucial role in understanding
effects of CP violation. Along the lines determined by (A) and (B) the
probabilities $P_{\mu e}$, $P_{e\mu}$ $P_{\tau e}$ and $P_{e\tau}$ do
not depend on the CP-phase. The other probabilities (as we will
discuss later) only weakly depend on the phase along these lines. The
lines shown in Fig.~\ref{fig:dcp-me} were calculated in the
factorization approximation, without assuming constant-density matter,
by solving numerically the corresponding 2-flavor evolution problems
with the PREM Earth density profile.

In what follows we will consider these lines and their connection to
the conditions in Eqs.~\eqref{eq:abc} in turn.

\subsubsection*{$\bullet$ Solar magic lines}

Let us discuss the condition $A_{S} (E_\nu, \Theta_\nu) = 0$. Notice
that at $A_{e\tilde{2}} \approx A_{S} = 0$ the ``solar'' contribution
to the amplitudes of the $\nu_\mu\leftrightarrow \nu_e$ and
$\nu_\tau\leftrightarrow \nu_e$ transitions vanishes, and
\begin{equation}
    P_{\mu e} = P_{e\mu} = s_{23}^2 |A_{e\tilde{3}}|^2 \,,
    \qquad
    P_{\tau e} = P_{e\tau} = c_{23}^2 |A_{e\tilde{3}}|^2 \,.
\end{equation}

In Fig.~\ref{fig:dcp-me} the condition $A_{S} = 0$ determines nearly
vertical lines at the values of the nadir angle $\Theta_\nu \approx
54^\circ$, $30^\circ$ and $12^\circ$. This feature can be immediately
understood using the constant density approximation. Indeed, according
to \eqref{eq:2nu-ampl} the condition $A_{S} = 0$ is fulfilled when
\begin{equation}
    \label{eq:sin-ft}
    \sin \phi_{S} (E_\nu, \Theta_\nu) = 0 \,.
\end{equation}
As follows from \eqref{eq:o12}, Eq.~\eqref{eq:sin-ft} is satisfied
when
\begin{equation}
    \label{eq:cond1}
    L(\Theta_\nu) \approx
    \frac{4E_\nu \pi n}{\Dmq_{21}\sqrt{(\cos 2\theta_{12}
	\mp 2 V E_\nu/\Dmq_{21})^2 + \sin^2 2\theta_{12}}} \,,
    \qquad n = 1,\, 2,\, \dots
\end{equation}
Furthermore, at energies that are much higher than the solar MSW
resonance energies in the mantle and in the core of the Earth,
$E_\nu\gtrsim 0.5~\GeV$, the condition \eqref{eq:cond1} becomes
\begin{equation}
    \label{eq:cond1a}
    L(\Theta_\nu)  \simeq \frac{2\pi n}{V} \,.
\end{equation}
Note that it is energy independent and determines the baselines for
which the ``solar'' contribution to the probability vanishes.  In the
plane $(\Theta_\nu, E_\nu)$ it represents nearly vertical lines
$\Theta_\nu \approx \text{const}$.

There are three solar magic lines which correspond to $n = 1$ (in the
mantle domain) and $n = 2, 3$ (in the core domain). The existence of a
baseline ($L\approx 7600$ km) for which the probability of
$\nu_e\leftrightarrow \nu_\mu$ oscillations in the Earth is
approximately independent of the ``solar'' parameters ($\Dmq_{21}$,
$\theta_{12}$) and of the CP-phase $\delta$ was first pointed out
in~\cite{Barger:2001yr} and later discussed in a number of
publications (see, \eg, Refs.~\cite{Huber:2002uy, Huber:2003ak,
Gandhi:2004bj, Blondel:2006su, Huber:2006wb, Smirnov:2006sm,
Agarwalla:2007ai}). This baseline was dubbed ``magic''
in~\cite{Huber:2002uy}.  The interpretation of this baseline as
corresponding to vanishing ``solar'' amplitude $A_{e\tilde{2}}$,
according to Eq.~\eqref{eq:cond1a} with $n=1$, was given
in~\cite{Smirnov:2006sm}. In~\cite{Smirnov:2006sm} it was also shown
that for neutrino trajectories crossing the core of the Earth there
exist two more solar ``magic'' baselines, corresponding to the
oscillation phase equal $\pi n $ with $n=2$ and 3, and the existence
of the atmospheric ``magic curves'' was pointed out. The three solar
``magic'' baselines, for which the amplitude $A_{e\tilde{2}}$
vanishes, can be clearly seen in the left panels of
Fig.~\ref{fig:solar}.

In the 1-3 resonance region and above, the factorization approximation
becomes invalid since the angle $\theta_{13}^m$ is large. As a result,
$A_{e\tilde{2}}$ and $A_{S}$ become substantially different. Indeed in
the constant density approximation, Eq.~\eqref{eq:ample2}, the
equality $A_{e\tilde{2}} = 0$ is satisfied when
\begin{equation}
    \label{eq:sin-ft1}
    \sin\phi_{21}^m (E_\nu, \Theta_\nu) = 0 \,.
\end{equation}
For $E_\nu \ll E_{13}^R$ we have $\phi_{21}^m \approx \phi_S$. But for
energies of the 1-3 resonance and above $\phi_{21}^m \neq \phi_S$. In
particular, for energies substantially above the resonance energy,
$\phi_{21}^m \approx \phi_A^0$, and according to
Eq.~\eqref{eq:phi21above},
\begin{equation}
    L(\Theta_\nu) \approx \frac{4E_\nu \pi n}{\Dmq_{31}} \,.
\end{equation}
Thus, in the 1-3 resonance region, the condition $\sin\phi_S \approx
0$ transforms into $\sin\phi_A^0 = 0$, and the lines of condition
$A_{e\tilde{2}} \approx 0$ bend. This can be seen in
Figs.~\ref{fig:dcp-me}, where the solar magic lines substantially
deviate from the lines of $\Delta P_{\mu e}^\text{CP} = 0$.

In the antineutrino channel no level crossing occurs and $\phi_{21}^m
\approx \phi_S$ everywhere.

\subsubsection*{$\bullet$ Atmospheric magic lines}

The atmospheric magic lines are determined by the condition $A_A
(E_\nu, \Theta_\nu) = 0$. When the condition $A_{e\tilde{3}}(E_\nu,
\Theta_\nu) \approx A_A (E_\nu, \Theta_\nu) = 0$ is satisfied, the
``atmospheric'' contribution to the amplitudes of $\nu_\mu
\leftrightarrow \nu_e$ and $\nu_\tau \leftrightarrow \nu_e$
transitions vanishes. In this case, too, there are no effects of CP
phase on the probabilities of oscillations involving $\nu_e$ or
$\bar{\nu}_e$.

The properties of atmospheric magic lines can be easily understood in
the constant density approximation. As follows from
Eq.~\eqref{eq:2nu-ampl}, the condition $A_A = 0$ is satisfied when
$\sin \phi_A = 0$ ($\phi_A = \pi k$, $k = 1, 2, \dots$) or explicitly
\begin{equation}
    \label{eq:cond2}
    L(\Theta_\nu) \approx \frac{4E_\nu \pi k}{\Dmq_{31}
      \sqrt{(\cos 2\theta_{13} \mp 2 V E_\nu/\Dmq_{31})^2
	+ \sin^2 2\theta_{13}}} \,,
    \qquad k = 1,\, 2,\, \dots
\end{equation}
For energies which are not too close to the atmospheric MSW resonance
energy, the condition \eqref{eq:cond2} reduces to
\begin{equation}
    \label{eq:cond2a}
    E_\nu \simeq \frac{\Dmq_{31} L(\Theta_\nu)}
    {|4\pi k \pm 2 V L(\Theta_\nu) |} \,,
\end{equation}
which corresponds to the bent curves in the $(\Theta_\nu, E_\nu)$
plane. For very large energies, where $\Dmq_{31}/2E \ll V$, the
atmospheric lines approach the same vertical lines as the solar magic
lines \eqref{eq:cond1a}.

Let us now consider the condition $A_{e\tilde{3}}(E_\nu, \Theta_\nu) =
0$. In the constant density approximation it gives, according to
\eqref{eq:ample3},
\begin{equation}
    \label{eq:condAaa}
    \sin \phi_{32}^m =
    -e^{i\phi_{31}^m} \cos^2 \theta_{12}^m \sin\phi_{21}^m \,.
\end{equation}
In turn, Eq.~\eqref{eq:condAaa} implies two conditions which follow
from the real and imaginary parts of the equality:
\begin{equation}
    \sin \phi_{32}^m
    = - \cos \phi_{31}^m \cos^2 \theta_{12}^m \sin\phi_{21}^m \,,
\end{equation}
\begin{equation}
    \sin \phi_{31}^m \sin\phi_{21}^m = 0.
\end{equation}
Both conditions can be satisfied simultaneously only at certain points
of the parameter space, which illustrates our general statement in the
beginning of this subsection. However at high energies $\cos^2
\theta_{12}^m \ll 1$, and the equality \eqref{eq:condAaa}
approximately reduces to
\begin{equation}
    \label{eq:magicatm}
    \sin \phi_{32}^m = 0 \,.
\end{equation}
Furthermore, since $\phi_{32}^m \approx \phi_A$ in the energy range
above the 1-2 resonance, in this channel the factorization
approximation works well.  The atmospheric magic lines reproduce very
well the lines of $\Delta P_{\mu e}^\text{CP} = 0$. No interconnection
of the atmospheric magic lines occurs.

\subsubsection*{$\bullet$ The interference phase condition}

Let us now find the lines in the ($\Theta_\nu, E_\nu$) plane which
correspond to the interference phase condition \eqref{eq:C}. We shall
call these lines the interference phase lines. Consider the condition
\eqref{eq:C} in the factorization approximation.  Although
$A_{e\tilde{2}}$ and $A_{e\tilde{3}}$ are pure imaginary in the bases
where their respective $2 \times 2$ Hamiltonians are traceless, their
relative complex phase $\phi$ in any fixed basis is different from
zero. It just equals the rotation phase between the aforementioned
2-flavor bases. In the constant density approximation
\eqref{eq:relphase} $\phi \approx - \phi_{31}^m$. In the energy range
between the two resonances we have
\begin{equation}
    \label{eq:betw}
    \phi_{31}^m \approx \frac{\Dmq_{31} L}{4E_\nu} = \phi_A^0 \,,
\end{equation}
\ie, in the first approximation $\phi$ does not depend on the matter
density. From \eqref{eq:C} we then obtain
\begin{equation}
    \frac{\Dmq_{31} L}{4E_\nu}
    = - \frac{\delta + \delta_0}{2} + \pi l \,,
\end{equation}
or
\begin{equation}
    \label{eq:intphase}
    E_\nu = \frac{\Dmq_{31} L(\Theta_\nu)}
    {4\pi l - 2(\delta + \delta_0)} \,.
\end{equation}
This gives rather accurate description of the lines $\Delta P_{\mu
e}^\text{CP} = 0$ below the 1-3 resonance.

\FIGURE[!t]{
  \includegraphics[width=141mm]{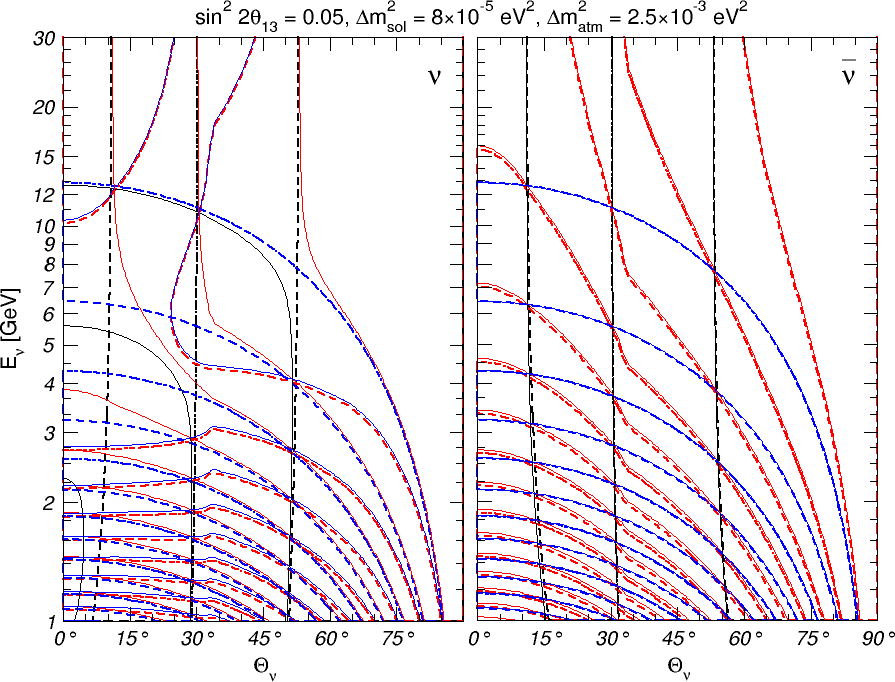}
  \caption{\label{fig:grid}%
    Grids of lines along which $\sin\phi_i = 0$ for different
    adiabatic phases $\phi_i$. Dashed lines correspond to $\phi_S$
    (black), $\phi_A$ (red) and $\phi_A^0$ (blue) obtained in $2\nu$
    context. Solid lines correspond to $3\nu$ calculations:
    $\phi_{21}^m$ (black), $\phi_{31}^m$ (red) and $\phi_{32}^m$
    (blue).}}

In the $3\nu$ framework $\phi_{31}^m$ differs from $\phi_A^0$ in the
1-3 resonance region.  Above the 1-3 resonance $\phi_{31}^m \approx
\phi_S$, and as we discussed before, the lines of $\phi_S =
\text{const}$ (Eq.~\eqref{eq:cond1a}) depend on energy. Hence, the
interference phase lines become nearly vertical with increasing
energy, as can be seen in Fig.~\ref{fig:grid}.

Summarizing, the phase $\phi_{21}^m$ that enters into the amplitude
$A_{e\tilde{2}}$ nearly coincides with $\phi_S$ below the 1-3
resonance. However, above the 1-3 resonance it approaches the vacuum
phase $\phi_A^0$. In turn, the interference phase $\phi \approx
\phi_{31}^m$, which approximately coincides with $\phi_A^0$ between
the 1-2 and 1-3 resonances, approaches $\phi_S$ with increase of
energy above the 1-3 resonance. The lines $\sin \phi_S = 0$ and $\sin
\phi_A^0 = 0$ cross, whereas $\sin \phi_{21}^m = 0$ and $\sin
\phi_{31}^m = 0$ do not. Thus, in the region of the 1-3 resonance,
where the factorization approximation is strongly broken in the 1-2
channel, interconnections of the lines occur: the contours of zero
$\Delta P_{\mu e}^\text{CP}$ transform (interpolate) from the solar
magic lines to the interference phase lines and vice versa. The
interconnection is related to the level crossing phenomenon and
reflects the level crossing scheme, it reflects the described change
of the phases $\phi_{21}^m$ and $\phi_{31}^m$.  To illustrate this
effect explicitly, we show in Fig.~\ref{fig:grid} the lines $\sin
\phi_S = 0$ and $\sin \phi_A = 0$, which represent the magic lines
(defined in the $2\nu$ context), and the lines $\sin \phi_{21}^m = 0$,
$\sin \phi_{31}^m = 0$ and $\sin \phi_{32}^m = 0$ which represent the
conditions of vanishing amplitudes in the $3\nu$ context, that is, the
contours of vanishing $\Delta P_{\mu e}^\text{CP}$. Clearly, they do
not coincide with the contours $\Delta P_{\mu e}^\text{CP} = 0$, since
we have taken into account the phase factors only. This grid
reproduces qualitatively well all the features of the lines shown in
Fig.~\ref{fig:dcp-me}.

There is no level crossing in the antineutrino channels for the normal
mass hierarchy, and therefore there is no interconnection of the lines
there.


\subsection{CP-phase domains for channels involving $\nu_e$}

The solar and atmospheric ``magic'' lines and the interference phase
curves allow a simple interpretation of the CP oscillograms. The solar
(nearly vertical) and atmospheric (bent) curves divide the
oscillograms into a set of domains, which are in turn divided by the
grid of the interference phase curves into sub-domains (see
Fig.~\ref{fig:dcp-me}). From these figures one can see that the
interference phase curves are steeper than the atmospheric curves in
the case of neutrinos and less steep than atmospheric curves for
antineutrinos, in full agreement with Eqs.~\eqref{eq:cond2a} and
\eqref{eq:intphase}. This fact is related with the sign in
Eq.~\eqref{eq:cond2a}. The probability difference $\Delta P_{\mu
e}^\text{CP}(\delta)$ vanishes at the borders of these sub-domains: On
the solar and atmospheric ``magic'' curves because the probabilities
are $\delta$-independent there, and on the interference phase curves
because they correspond to $\cos(\phi - \delta) = \cos(\phi -
\delta_0$).
The signs of the probability differences in the neighboring
sub-domains are opposite, with the maxima of the difference in the
central parts of sub-domains.

As can be seen from Fig.~\ref{fig:dcp-me}, with changing the true (or
assumed) values of $\delta$, the solar and atmospheric grids remain
unchanged, whereas the grid of the interference phase curves moves up
or down, in accord with Eqs.~\eqref{eq:cond1a}, \eqref{eq:cond2a} and
\eqref{eq:intphase}. The constant-density approximation results of
Eqs.~\eqref{eq:cond1a}, \eqref{eq:cond2a} and \eqref{eq:intphase}
reproduce the main features of these curves quite well.

As can be seen from the figures, the borders between the regions of
the positive and negative CP-phase effect do not coincide exactly with
the magic lines, especially in the regions of intersection of these
lines. This indicates deviation from the factorization approximation
and is related to the level crossing phenomenon, as we have discussed
in the previous subsection.


\subsection{CP-domains for channels not involving $\nu_e$}

\PAGEFIGURE{
  \includegraphics[width=143mm]{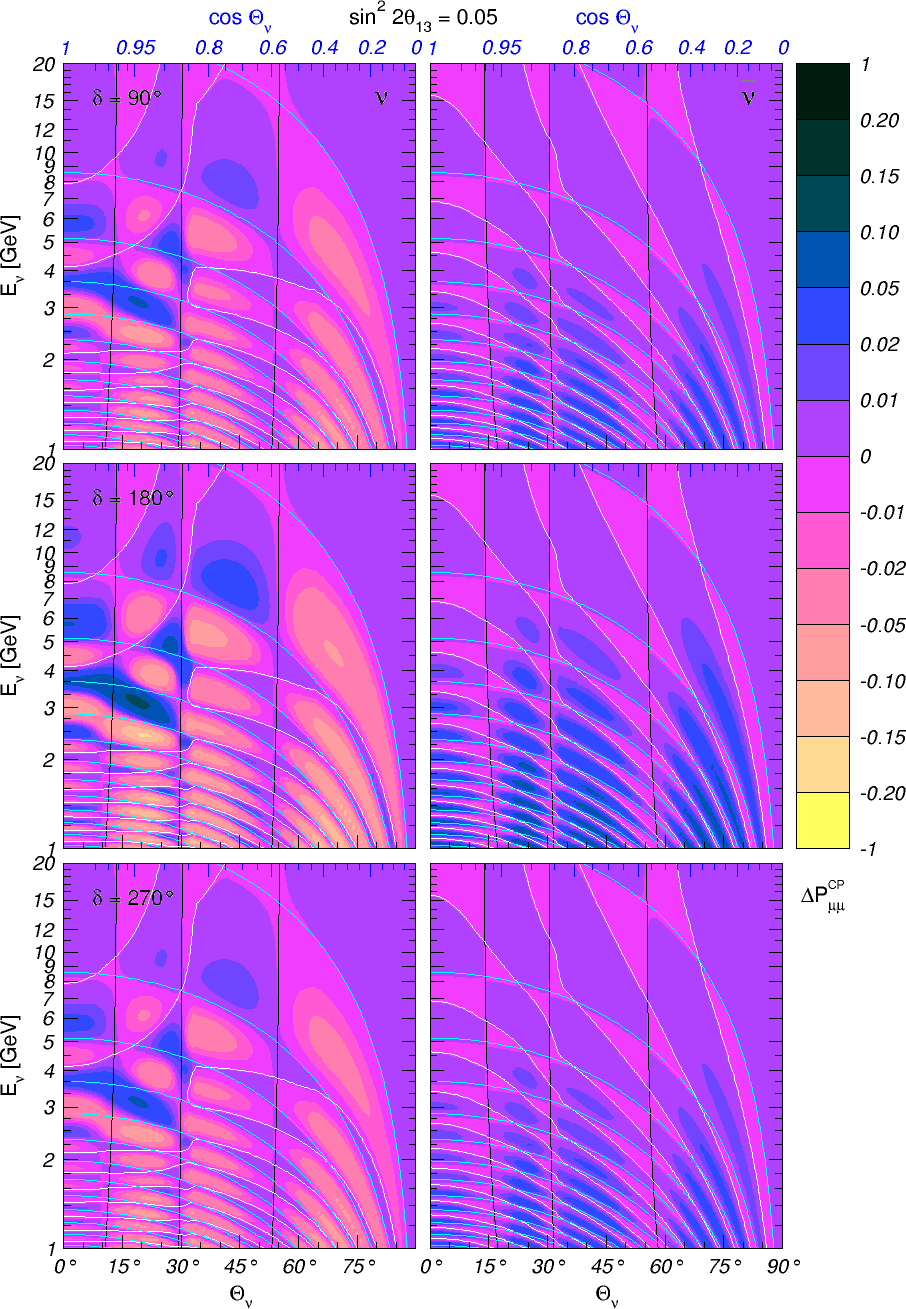}
  \caption{\label{fig:dcp-mm}%
    Oscillograms for the difference of probabilities $\Delta
    P_{\mu\mu}^\text{CP}(\delta) = P_{\mu\mu}(\delta) -
    P_{\mu\mu}(\delta_0)$ with $\delta_0 = 0^\circ$. Shown are the
    solar (black), atmospheric (white) and interference phase
    condition (cyan) curves.}}

As follows from our consideration in Sec.~\ref{sec:interf}, the
$\delta$-dependent parts of the probabilities for the channels which
do not contain $\nu_e$ have more complicated structure than those
which do. Apart from the term proportional the product of the
amplitudes $A_{e\tilde{2}} A_{e\tilde{3}}^*$ they contain
contributions proportional to the deviation of the 2-3 mixing from the
maximal one as well as terms $P^{\delta\delta}$ proportional to the
square of the small amplitude $A_{\tilde{2}\tilde{3}}$. In what
follows we will neglect the latter.

In the case of the maximal 2-3 mixing the $\delta$-dependent parts of
the probabilities are given in Eqs.~\eqref{eq:mumumax}
and~\eqref{eq:mutaumax}. Furthermore, $P_{\tau\tau}^\delta =
-P_{\mu\mu}^\delta$ and $P_{\tau\mu}^\delta = -P_{\mu\tau}^\delta$.
From Eq.~\eqref{eq:mumumax} it follows that for $\theta_{23} =
45^\circ$ the CP-oscillograms for the survival probability
$P_{\mu\mu}$ can be interpreted in terms of the same grids of the
solar and atmospheric ``magic'' curves $A_S = 0$ and $A_A = 0$ that we
used for the analysis of the oscillograms for $P_{\mu e}$ and $P_{\tau
e}$ (see Fig.~\ref{fig:dcp-mm}).

The key difference from the previous case is that now the CP-phase and
the interference phase dependencies factorize. In the interference
terms $P_{\mu\mu}^\delta$ and $P_{\tau\tau}^\delta$ they appear as
$\cos\delta \cos\phi$. Therefore, the third grid describing
oscillograms in these channels consists of the interference phase
curves
\begin{equation}
    \label{eq:phase1}
    \phi \approx - \phi_{31}^m = \frac{\pi}{2}+\pi n \,,
\end{equation}
which, unlike the interference phase curves for the $\nu_\mu - \nu_e$
and $\nu_\tau - \nu_e$ channels, do not depend on the values of
$\delta$ and $\delta_0$. The difference between the probabilities
calculated with the true and assumed values of $\delta$ (as well as
the effects of $\delta$ in general) vanishes on the curves of all
three types and take maximum values in the central parts the domains
delimited by these curves. The borders of the domains do not move with
change of $\delta$, and the only change that happens is that within
each domain the probability varies proportionally to $\cos\delta$.

The $\delta$ dependent terms of the transition probabilities,
$P_{\mu\tau}^\delta$ and $P_{\tau\mu}^\delta$ are proportional to
$\sin \delta \sin \phi$. Therefore for these probabilities the
interference phase condition reads
\begin{equation}
    \label{eq:phasemutau}
    \phi \approx \phi_{31}^m = \pi n \,,
\end{equation}
Again the borders of the domains do not depend on $\delta$ and with
changing $\delta$ the interference terms vary as $\sin\delta$.

\FIGURE[!t]{
  \includegraphics[width=143mm]{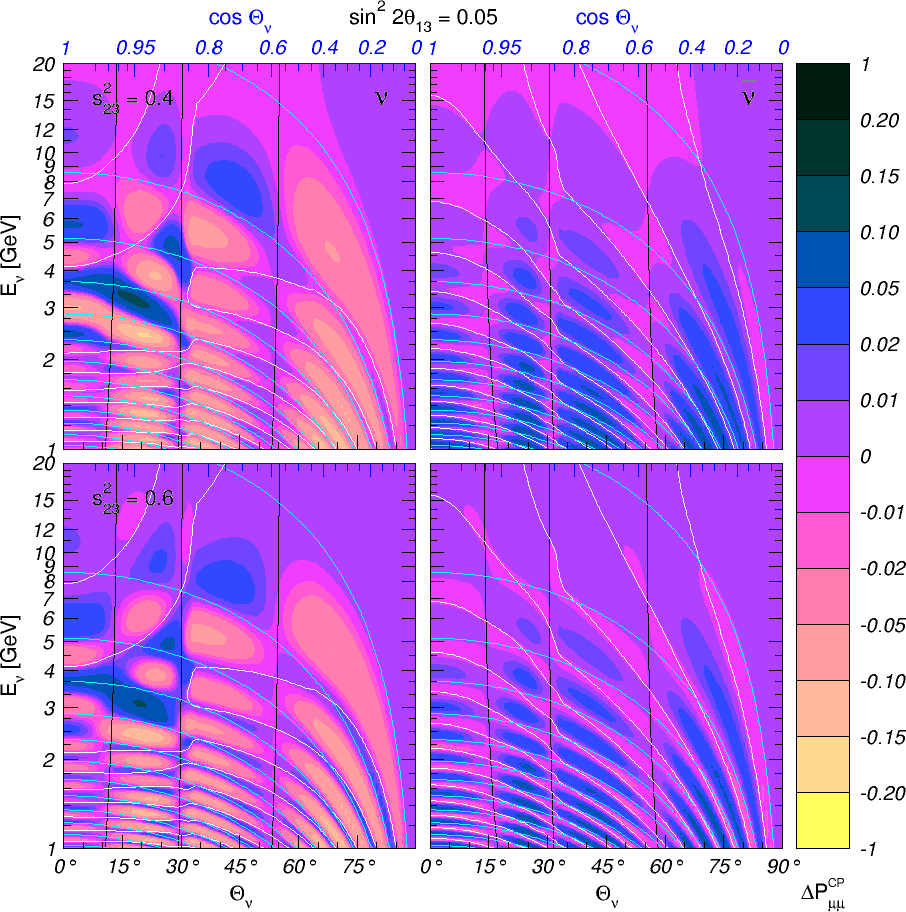}
  \caption{\label{fig:dcp-nomax}%
    Contour plots for the probability difference $\Delta
    P_{\mu\mu}^\text{CP}(\delta) = P_{\mu\mu}(\delta) -
    P_{\mu\mu}(\delta_0)$ with $\delta_0 = 0^\circ$ and $\delta =
    180^\circ$.  Upper panels: $s_{23}^2 = 0.4$, lower panels:
    $s_{23}^2 = 0.6$. Shown are the solar (black), atmospheric (white)
    and interference phase (cyan) curves.}}

Notice that the borders of domains are rather stable with respect to
variations of the neutrino parameters. Since the grids are determined 
mainly by the phases, their dependence on the 1-3 mixing is weak. With
the decrease of the 1-3 mixing angle in vacuum the grid lines becomes
closer to the lines of vanishing $\Delta P_{\mu e}^\text{CP}$. Since
$\theta_{23}$ is experimentally known to be rather close to
$45^\circ$, the discussed solar, atmospheric and interference phase
curves give a rather good description of the CP-oscillograms for
$P_{\mu\mu}$ even when $\theta_{23}$ deviates from the maximal-mixing
value (see Fig.~\ref{fig:dcp-nomax}).\footnote{Actually, the
correction due to $\theta_{23} \ne 45^\circ$ is of the order
$\frac{1}{2} \cos2\theta_{23} \lesssim 0.15$.}


\subsection{Sensitivity to the CP phase in the $\nu_\mu - \nu_e$ channel}
\label{sec:sens-me}

Let us now identify the regions of the experimental parameters
$\Theta_\nu$ and $E_\nu$ for which the oscillation probabilities have
maximal sensitivity to the CP phase $\delta$.

Consider the variation of the oscillation probabilities with varying
$\delta$ while all the other oscillation parameters are fixed. As
follows from Eq.~\eqref{eq:Pme-2}, the maximal variation of the
probability $P_{\mu e}$ with $\delta$ changing between $0^\circ$ and
$360^\circ$ is
\begin{equation}
    \label{eq:DP1}
    \Delta P_{\mu e}^\text{max} \equiv
    \max[ P_{\mu e}(\delta) ] - \min[ P_{\mu e}(\delta) ] =
    2 \sin 2 \theta_{23}\, |A_{e\tilde{2}} A_{e\tilde{3}}| \,.
\end{equation}
(Note that a similar quantity was considered in~\cite{Kimura:2006jj}).
This quantity is maximized when $|A_{e\tilde{2}} A_{e\tilde{3}}|$
takes the maximum possible value. Let us now discuss the dependence of
$|A_{e\tilde{2}} A_{e\tilde{3}}|$ on the experimental parameters
$\Theta_\nu$ and $E_\nu$ in the factorization approximation.
For mantle-only crossing trajectories it is sufficient to use the
constant-density matter factorization approximation, in which
\begin{equation}
    |A_{e\tilde{2}}| = \sin 2\theta_{12}^m \,|\sin\phi_S| \,, \qquad
    |A_{e\tilde{3}}| = \sin 2\theta_{13}^m \,|\sin\phi_A| \,.
\end{equation}
Thus, we have to find the maxima of the quantity
\begin{equation}
    \label{eq:A}
    A\equiv \sin 2\theta_{12}^m\, \sin 2\theta_{13}^m \,
    |\sin\phi_{21}^m \, \sin\phi_{31}^m| \,
\end{equation}
with respect to $\Theta_\nu$ and $E_\nu$. To do this exactly is a
rather complicated problem, and the result would be bulky and not
easily tractable; fortunately, an approximate maximization can be
readily carried out by studying the energy dependence of the different
factors in \eqref{eq:A}.

First, recall that for energies $E_\nu \gtrsim 0.5~\GeV$ the ``solar''
phase $\phi_S$ is essentially energy independent. Therefore $\sin
\phi_S$ can be considered a constant factor when maximizing $A$ with
respect to the neutrino energy. The maximum $|\sin \phi_S|=1$ can be
achieved by properly choosing the values of the baseline $L$. These
are approximately equal to the central values of $L$ (\ie, of
$\cos\Theta_\nu$) between the solar ``magic'' lines in the
oscillograms of Fig.~\ref{fig:dcp-me} (note that in the second band,
due to the existence of the mantle-core boundary, the maximum is
shifted from the center).

Next, consider the extrema of the remaining factor, $\sin
2\theta_{12}^m\, \sin 2\theta_{13}^m \,|\sin\phi_A|$, with respect to
the neutrino energy for fixed $L$. To do this, we make use of the fact
that the function
\begin{equation}
    \label{eq:f}
    f(E_\nu) \equiv \sin 2\theta_{12}^m\, \sin 2\theta_{13}^m
\end{equation}
varies with $E_\nu$ significantly more slowly than $\sin\phi_A$. If
$f(E_\nu)$ were constant, the maxima of $A$ with respect to $E_\nu$
would coincide with the maxima of $|\sin\phi_A|$; in reality, the
(relatively) weak energy dependence of $f(E_\nu)$ on $E_\nu$ leads to
a slight shift of the exact maxima $A$ from those of $|\sin\phi_A|$.
The main effect of the energy dependence of $f(E_\nu)$ is actually to
modulate the maxima of $|\sin\phi_A|$. In other words, the maxima of
$\Delta P_{\mu e}^\text{max}$ nearly coincide with the absolute maxima
of $|\sin\phi_S \sin\phi_A|$ (equal to 1), which are achieved by a
proper choice of the values of the nadir angle and energy; the actual
height of the local maxima of $\Delta P_{\mu e}^\text{max}$ is
determined by the value of $f(E_\nu)$ in these maxima.

\FIGURE[!t]{
  \includegraphics[width=142mm]{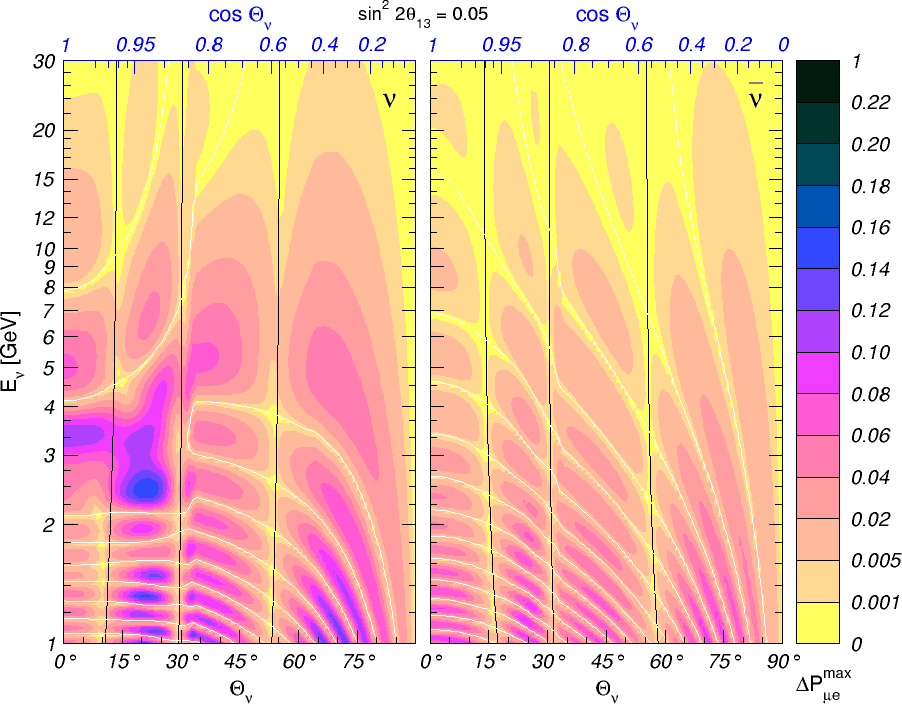}
  \caption{\label{fig:dmax-me}%
    Contour plots for the probability difference $\Delta P_{\mu
    e}^\text{max} = \max P_{\mu e} - \min P_{\mu e}$ for $\delta$
    varying between $0^\circ$ and $360^\circ$ and all the other
    oscillation parameters fixed. Also shown are the solar (black) and
    atmospheric (white) magic curves. We set $\Dmq_{21} = 8 \times
    10^{-5}~\eVq$ and $\sin^2 2\theta_{13} = 0.05$.}}

This is illustrated by Fig.~\ref{fig:dmax-me}. In the $\Theta_\nu$
direction, the maxima of $\Delta P_{\mu e}^\text{max}$ correspond to
the maxima of $|\sin \phi_S|$. This fixes the values of the baseline
$L$. The vertical ``domain structure'' (\ie, the structure in the
energy direction) is due to the oscillatory dependence of $|\sin
\phi_A|$. For fixed $L$ the energies $E_n$ at which $\Delta P_{\mu
e}^\text{max}$ has maxima are approximately found from the condition
$|\sin \phi_A| = 1$, or $\phi_A\approx \omega_{31} L = \pi/2 + \pi n$.
The heights of these local maxima are determined by $f(E_n)$.

To find out which of the peaks of $\Delta P_{\mu e}^\text{max}$ are
the highest, one has to consider the maxima of the envelop function
$f(E_\nu)$. We will do that here for the resonance channel
(neutrinos); the analysis for antineutrinos can be easily performed
along the same lines.

The condition $df / dE_\nu = 0$ yields the third-order equation
\begin{multline}
    \label{eq:char}
    2 x^3 - 3 x^2 (\cos2\theta_{13} + r_\Delta \cos2\theta_{12})
    \\
    + x(1 + 4r_\Delta \cos2\theta_{12} \cos2\theta_{13} + r_\Delta^2)
    - r_\Delta \cos2\theta_{12} - r_\Delta^2 \cos2\theta_{13} = 0,
\end{multline}
where $x$ is defined in \eqref{eq:def-x}. For $\sin^2 2\theta_{13}
\lesssim 1/9$ it has three solutions, which correspond to two maxima
of $f(E_\nu)$ and a minimum between the two maxima. Expanding in
$r_\Delta$, we find for the position of the low-energy maximum
\begin{equation}
    x_1 \simeq r_\Delta \cos 2\theta_{12} +
    r^2_\Delta \sin^2 2\theta_{12} \cos 2\theta_{13} \,,
\end{equation}
or
\begin{equation}
    E_1 \simeq E_{12}^R\, (1 +
    r_\Delta \sin 2\theta_{12}\tan 2\theta_{12} \cos 2\theta_{13}) \,.
\end{equation}
Here $E_{12}^R$ is the energy of the 1-2 resonance (which corresponds
to $x = r_\Delta \cos 2\theta_{12}$). Thus, the low-energy maximum of
$f(E_\nu)$ practically coincides with the low-energy MSW resonance,
and the 1-3 mixing produces only a slight ($\mathcal{O}(r_\Delta)$)
upward shift of the position of the maximum.

The minimum of $f(E_\nu)$ is given by
\begin{equation}
    x_2 \simeq \frac{3}{4} \cos 2\theta_{13} -
    \frac{1}{4}\sqrt{1 - 9 \sin^2 2\theta_{13}} \,,
\end{equation}
while the second maximum is at
\begin{equation}
    x_3 \simeq \frac{3}{4} \cos2\theta_{13} +
    \frac{1}{4}\sqrt{1 - 9 \sin^2 2\theta_{13}} \,.
\end{equation}
Recall that the MSW resonance energy in the 1-3 channel corresponds to
$x = x_{13}^R \equiv \cos 2\theta_{13}$.

Consider the dependence of $f(E_\nu)$ on $E_\nu$ for different values
of the 1-3 mixing. For $\theta_{13} \to 0$ we have $x_3 \to 1$ and
$x_2 \to 1/2$, \ie, $E_3 \to E_{13}^R$ and $E_2 \to E_{13}^R / 2$.
Thus, in this limit the positions of the maxima of $f(E_\nu)$ coincide
with the 1-2 and 1-3 MSW resonance energies, while the minimum is
approximately in the middle between them.

At maxima, the values of function $f$ can be estimated as $f(x_1) \sim
\sin 2 \theta_{13}$ and $f(x_1) \sim r_\Delta \sin 2 \theta_{12}$.

With increasing 1-3 mixing, the minimum of $f(E_\nu)$ shifts to larger
energies:
\begin{equation}
    x_2 \approx \frac{1}{2} + \frac{3}{4} \sin^2 2\theta_{13} \,,
\end{equation}
whereas the second maximum moves to lower energies:
\begin{equation}
    x_3 \approx 1 - \frac{3}{2}\sin^2 2\theta_{13}
    \approx x_{13}^R \cos^2 2\theta_{13} \,.
\end{equation}
For $\sin^2 2\theta_{13} = 1/9$ one has $x_2 = x_3 = 1/\sqrt{2}$, \ie,
the minimum and the second maximum of $f(E_\nu)$ merge. This just
corresponds to the situation when the local minimum of the cubic
function on the l.h.s. of Eq.~\eqref{eq:char} touches the $x$-axis.

For $\sin^2 2\theta_{13} > 1/9$ only the low-energy maximum of
$f(E_\nu)$ persists. Thus, for these values of $\sin^2 2\theta_{13}$
the effect of the CP phase is maximal at the 1-2 resonance and
decreases with increasing energy. For the other value, $\sin^2
2\theta_{13} = 0.05$ (which in fact is not too small), we find $x_2
\approx 0.57$ and $x_3 \approx 0.89$.

These results allow one to readily understand the oscillograms for
$\Delta P_{\mu e}^\text{max}$ (Fig.~\ref{fig:dmax-me}), at least for
mantle-only crossing neutrinos.  For $\sin^2 2\theta_{13} = 0.05$ the
minimum of $\Delta P_{\mu e}^\text{max}$ is situated at $E_\nu \sim 3
- 3.5~\GeV$ ($x \approx 0.57$), which can be seen in the strip between
the first and the second solar magic lines: the peaks increase in
height both with energies decreasing and increasing from $E_\nu \sim 3
- 3.5~\GeV$ (in contrast to this, for $\sin^2 2\theta_{13} = 0.125$
the height of the peaks would monotonically decrease with increasing
neutrino energy). Note that the situation is somewhat different in the
strip between the Earth's surface ($\Theta_\nu = 90^\circ$) and the
first solar ``magic'' line: the peak at $E_\nu\sim 5~\GeV$ is actually
lower than that at $\sim 3~\GeV$ because the baseline is relatively
short, and at high energies the oscillation phase $\phi_A$ is too
small for the condition $|\sin\phi_A| = 1$ to be satisfied.

For core-crossing neutrino trajectories the constant-density
approximation is not in general applicable, and a different approach
is necessary.  The maxima of $\Delta P_{\mu e}^\text{max}$ in that
case can, in principle, be analyzed in the factorization approximation
by making use of simple formulas for 2-flavor neutrino evolution in
3-layer matter density profiles obtained in~\cite{Akhmedov:1998ui}. In
Fig.~\ref{fig:dmax-me} one can see strong enhancement of the
difference of amplitudes in the core domain at $E_\nu \sim 2.5 $ GeV, 
which is apparently due to the mantle-core effect.


\subsection{Sensitivity to the CP phase in the $\nu_\mu - \nu_\mu$
  channel}

Let us now discuss the sensitivity of the survival probability
$P_{\mu\mu}$ to the phase $\delta$. For other discussions of this
issue see, \eg,~\cite{Kimura:2006hy, Kimura:2007mu, Kimura:2008nq}.
From Eq.~\eqref{eq:Pmumu} one finds
\begin{equation}
    \label{eq:Pmumu2}
    P_{\mu\mu} = |C + D\, z|^2 \,,
\end{equation}
where
\begin{equation}
    \label{eq:CD}
    C = c_{23}^2 A_{\tilde{2}\tilde{2}} + s_{23}^2 A_{\tilde{3}\tilde{3}}\,,\qquad
    D = 2\,s_{23}\,c_{23}\,A_{\tilde{2}\tilde{3}}\,, \qquad
    z \equiv \cos\delta\,.
\end{equation}
The maximum and minimum values of $P_{\mu\mu}$ with varying $\delta$
then correspond to the maximum and minimum of the modulus of the
complex number $C + D\, z$ when $C$ and $D$ are fixed and $z$ is
allowed to vary between -1 and 1. A simple geometrical consideration
then shows that for
\begin{equation}
    \label{eq:cond4}
    2\, s_{23}\, c_{23}\, |A_{\tilde{2}\tilde{3}}|^2 \le
    \LT| \Re[ A_{\tilde{2}\tilde{3}}^*
    (c_{23}^2 A_{\tilde{2}\tilde{2}}
    + s_{23}^2 A_{\tilde{3}\tilde{3}}) ] \RT|
\end{equation}
the minimum of $P_{\mu\mu}$ corresponds to $\delta = 0$ and maximum to
$\delta = 180^\circ$ or vice-versa, so that for $\Delta
P_{\mu\mu}^\text{max} \equiv \max[P_{\mu\mu}(\delta)] -
\min[P_{\mu\mu}(\delta)]$ one finds
\begin{equation}\begin{split}
    \label{eq:DP2}
    \Delta P_{\mu\mu}^\text{max} &=
    \LT| P_{\mu\mu}(\delta=180^\circ) - P_{\mu\mu}(\delta=0^\circ) \RT|
    \\[1mm]
    &= 8\, s_{23}\, c_{23} \LT| \Re[ A_{\tilde{2}\tilde{3}}^*
    (c_{23}^2 A_{\tilde{2}\tilde{2}} + s_{23}^2
    A_{\tilde{3}\tilde{3}})] \RT| \,.
\end{split}\end{equation}
If, on the contrary,
\begin{equation}
    \label{eq:cond5}
    2\, s_{23}\, c_{23}\, |A_{\tilde{2}\tilde{3}}|^2 >
    \LT| \Re[ A_{\tilde{2}\tilde{3}}^*(c_{23}^2 A_{\tilde{2}\tilde{2}}
    + s_{23}^2 A_{\tilde{3}\tilde{3}}) ] \RT| \,,
\end{equation}
then for the maximal variation of $P_{\mu\mu}$ with $\delta$ one finds
\begin{equation}
    \label{eq:DP3}
    \Delta P_{\mu\mu}^\text{max}
    = \LT( \frac{\big|\! \Re[A_{\tilde{2}\tilde{3}}^*
      (c_{23}^2 A_{\tilde{2}\tilde{2}} + s_{23}^2 A_{\tilde{3}\tilde{3}})]
      \big|}{|A_{\tilde{2}\tilde{3}}|}
    + 2\, s_{23}\, c_{23}\, |A_{\tilde{2}\tilde{3}}| \RT)^2 \,.
\end{equation}

\FIGURE[!t]{
  \includegraphics[width=142mm]{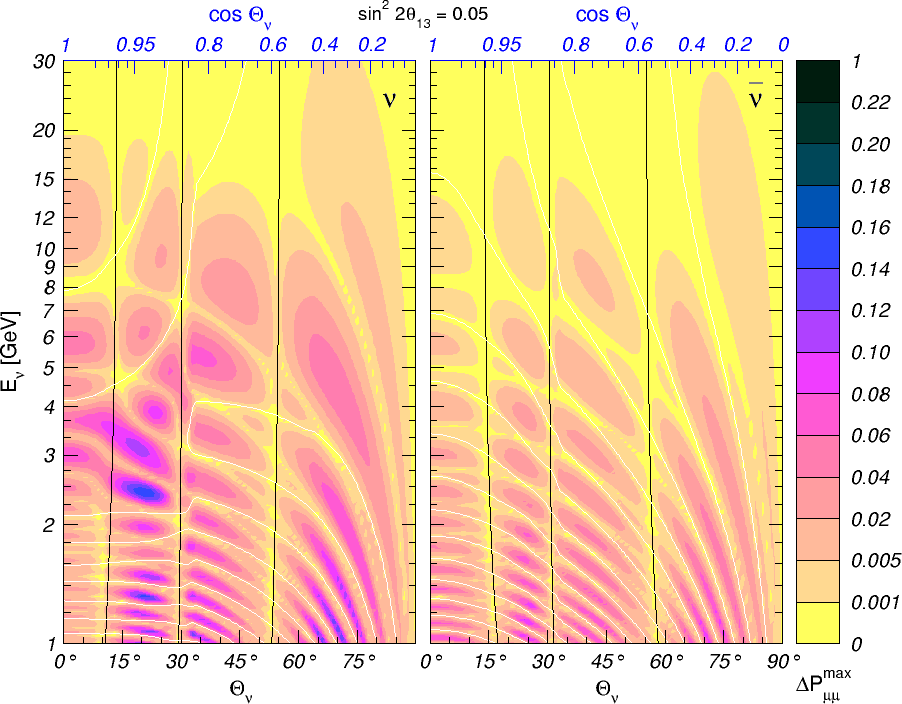}
  \caption{\label{fig:dmax-mm}%
    Contour plots for the probability difference $\Delta
    P_{\mu\mu}^\text{max} = \max P_{\mu\mu} - \min P_{\mu\mu}$ for
    $\delta$ varying between $0^\circ$ and $360^\circ$ and all the
    other oscillation parameters fixed. Also shown are the solar
    (black) and atmospheric (white) magic curves. We set $\Dmq_{21} =
    8 \times 10^{-5}~\eVq$ and $\sin^2 2\theta_{13} = 0.05$.}}

Note that $A_{\tilde{2}\tilde{3}}$ is a small quantity, so that the
condition \eqref{eq:cond4} is satisfied in most of the parameter
space. Exceptions are the regions where the ``main'' contribution to
$\Delta P_{\mu\mu}^\text{max}$, \ie, $|c_{23}^2 A_{\tilde{2}\tilde{2}}
+ s_{23}^2 A_{\tilde{3}\tilde{3}}|^2$, is anomalously small, that is
the regions along the magic lines. This is illustrated by
Fig.~\ref{fig:dmax-mm}, where we show the oscillograms for the maximum
probability differences $\Delta P_{\mu\mu}^\text{max}$ for $\delta$
varying between $0^\circ$ and $360^\circ$. The areas corresponding to
the regions where the condition \eqref{eq:cond4} is not satisfied
occupy a rather small fraction of the parameter space. Moreover,
$\Delta P_{\mu\mu}^\text{max}$ is small in these regions, so that they
correspond to low sensitivity to the effects of the CP phase. Note
that these regions would never appear if one neglected the term of the
order of $|A_{\tilde{2}\tilde{3}}|^2$ in the expression for
$P_{\mu\mu}$.

For the regions where the condition \eqref{eq:cond4} is satisfied, one
can obtain a simple expression for $\Delta P_{\mu\mu}^\text{max}$ in
the limit $\theta_{23}=45^\circ$. From Eqs.~\eqref{eq:DP2} and
\eqref{eq:unit} one then finds
\begin{equation}
    \label{eq:DP4}
    \Delta P_{\mu\mu}^\text{max}
    = 2\, |\Re(A_{e\tilde{2}} A_{e\tilde{3}}^*)| \,.
\end{equation}
Thus, for $\theta_{23} = 45^\circ$ the oscillograms for the maximum
probability difference $\Delta P_{\mu\mu}^\text{max}$ are also
governed by the solar and atmospheric ``magic'' curves, as well as by
the interference phase curves \eqref{eq:phase1}, in full accord with
our discussion in Sec.~\ref{sec:magic}.


\section{Discussion and conclusions}
\label{sec:concl}

The main purpose of the present paper is to gain a physics insight
into the complex pattern of full 3-flavor neutrino oscillations in the
Earth. To this end, we presented a detailed description of the
three-flavor neutrino oscillation effects in the Earth in terms of the
neutrino oscillograms, \ie, contours of equal oscillation
probabilities or probability differences in the neutrino nadir
angle~--~energy plane.

We have found that for very small or vanishing 1-3 mixing the 
oscillation pattern appears in the low energy region with large 
(maximal) transition probabilities below $0.3~\GeV$. In the mantle 
domain the oscillation pattern consists of three MSW resonance peaks,
which correspond to the oscillation phases $\pi/2$, $3\pi/2$ and
$5\pi/2$, and the parametric resonance ridge in the core domain at
$E_\nu \approx 0.2~\GeV$ and $\Theta_\nu \sim 28 - 30^{\circ}$. 

For non-zero 1-3 mixing the oscillograms consists of the low energy
pattern, where the effect of the 1-2 mixing dominates, and the high
energy pattern, determined mainly by the 1-3 mixing and mass
splitting, if the 1-3 mixing is not too small. The low energy pattern
is modulated by the high frequency and small amplitude effect induced
by the 1-3 mode, whereas the high energy structure is modulated by the
low (refraction) frequency small amplitude effect due to the 1-2 mass
splitting and mixing.

We studied in detail the effect of the 1-2 mode on the oscillograms
for energies $E_\nu > 1~\GeV$. At these energies, if $\theta_{13}$ is
not very small, the oscillation pattern is determined mainly by the
1-3 mixing and mass splitting, whereas the 1-2 mass splitting and
mixing lead to small corrections. In the $\nu_e - \nu_e$ channel the
interference of the 1-2 and 1-3 modes is strongly suppressed, and the
effect of the 1-2 mixing is due to corrections to the 1-3 mixing and
the atmospheric phase.  In the $\nu_\mu - \nu_e$ channel the effect of
the 1-2 mixing is dominated by the interference of the solar and
atmospheric amplitudes. These corrections have a domain structure in
the $E_\nu - \Theta_\nu$ plane. In the $\nu_\mu - \nu_\mu $ and
$\nu_\mu - \nu_\tau$ channels the effect of the 1-2 mixing is
essentially due to the corrections to the phase of the main (vacuum)
oscillation mode.
This consideration is important in discussions of the degeneracies of
parameters in terms of the oscillograms.

We studied the properties of the interference of the amplitudes
$A_{e\tilde{2}}$ and $A_{e\tilde{3}}$ and, in particular, the effects
of CP-violation which are associated with this term. The structure of
the interference term is simply and rather accurately determined in
the factorization approximation, when one has $A_{e\tilde{3}} \approx
A_A (\Dmq_{31}, \theta_{13})$ and $A_{e\tilde{2}} \approx A_S
(\Dmq_{21}, \theta_{12})$. This means that the dependence of the
parameters of 1-2 sector and 1-3 sector factorizes in the interference
term.  This approximation does not work in the resonance regions.

We showed that the interference term, and therefore CP-violation,
exhibit a domain structure in the $E_\nu - \Theta_\nu$ plane. The
borders of the domains are determined by the grids of magic lines
(solar and atmospheric) and by the lines of the interference phase
condition. In the neighboring domains the sign of the CP effect is
opposite. Beyond the factorization approximation the interconnections
of the solar and phase condition lines occur, which are related to the
level crossing phenomenon.

We studied the dependence of the probabilities on the CP-phase. The
character of the dependence on the CP-phase is different for survival
and transition probabilities.  In the standard parametrization, the
$\nu_e$ survival probability does not depend on $\delta$. For the
survival channels $\nu_\mu - \nu_\mu$ and $\nu_\tau - \nu_\tau$ as
well as for the transition channel $\nu_\mu - \nu_\tau$ all three sets
of the lines --~the borders of the domains~-- do not depend on
$\delta$. Within a given domain the interference term changes as
$\propto \cos \delta$ in the $\nu_\mu - \nu_\mu$ channel, and as
$\propto \sin \delta$ in the $\nu_\mu - \nu_\tau$ channel. For the
probabilities of transitions which involve $\nu_e$, \ie\ $\nu_e -
\nu_\mu$ and $\nu_e - \nu_\tau$, the interference phase condition
depends on $\delta$, so that with changing $\delta$ the corresponding
lines move. Therefore, the modification of the pattern of CP violation
is determined by this motion of the interference phase lines.

The phase $\delta$ can affect significantly all the oscillation
probabilities but $\nu_e\leftrightarrow \nu_e$. Thus, in principle,
one can study the effects of leptonic CP violation by precision
measurements of the energy and zenith angle dependence of these
probabilities. We find that the strongest variation of the probability
with $\delta$ occurs in the region of the 1-2 resonance. For $\sin^2
2\theta_{13} < 1/9$ the local maximum appears in the 1-3 resonance
region. The weakest variation is at $E_\nu \approx 0.5 E_{13}^R$.

Many features of the oscillograms discussed in this paper are 
unobservable in the present and forthcoming experiments. The 
accelerator experiments cover only several peripheral regions of the
oscillograms, which correspond to large values of the nadir angles
$\Theta_\nu > 77^{\circ}$. The large underwater and ice detectors have
high energy thresholds $E > 15$ GeV. Thus, the most interesting and
structured regions of the oscillograms turn out to be uncovered.
Measurements of the oscillograms with atmospheric neutrinos could be
performed using multi-megaton water Cherenkov detectors, which will
have sufficient statistics in the energy range $E >$ 1 -- 2 GeV, where
the energy and angular resolutions are good
enough~\cite{Suzuki:2006bq, ournew}. Detailed consideration of various
features of the oscillograms performed in this paper can help develop
methods which will enhance the sensitivity of future experiments to
the CP-violating phase and other neutrino parameters.


\acknowledgments

A.Yu.S.\ is grateful to the Max-Planck-Institut f\"ur Kernphysik,
Heidelberg, for hospitality. The work of A.Yu.S.\ has been supported
in part by the Alexander von Humboldt Foundation.
M.M.\ is supported by MCI through the Ram\'on y Cajal program and
through the national project FPA2006-01105, and by the Comunidad
Aut\'onoma de Madrid through the HEPHACOS project P-ESP-00346.


\providecommand{\href}[2]{#2}\begingroup\raggedright\endgroup

\end{document}